\documentclass[a4paper,11pt]{article}
\usepackage{bm,amsmath,amssymb,mathrsfs,graphicx,citesort}
\newcommand{\nc}{\newcommand}
\nc{\rnc}{\renewcommand}
\nc{\nn}{\nonumber}
\rnc{\Im}{{\rm{Im}\,}}
\rnc{\Re}{{\rm{Re}\,}}
\rnc{\i}{{\rm i}}
\rnc{\d}{{\rm d}}
\nc{\e}{{\rm e}}
\nc{\mfa}{{\mathfrak{a}}}
\nc{\mfab}{\overline{\mfa}}
\nc{\mfA}{{\mathfrak{A}}}
\nc{\mfAb}{\overline{\mfA}}
\nc{\mfb}{{\mathfrak{b}}}
\nc{\mfbb}{\overline{\mfb}}
\nc{\mfB}{{\mathfrak{B}}}
\nc{\mfBb}{\overline{\mfB}}
\nc{\mfc}{{\mathfrak{c}}}
\nc{\mfcb}{\overline{\mfc}}
\nc{\mfC}{{\mathfrak{C}}}
\nc{\mfCb}{\overline{\mfC}}
\nc{\db}{\displaybreak[0]\\}
\nc{\bra}{\langle}
\nc{\ket}{\rangle}
\nc{\bbra}{\left\langle}
\nc{\kket}{\right\rangle}
\nc{\J}{\mathscr{J}}
\nc{\R}{\mathscr{R}}
\nc{\T}{\mathscr{T}}
\rnc{\H}{\mathscr{H}}
\nc{\M}{\mathscr{M}}
\nc{\Q}{\mathscr{Q}}
\nc{\A}{\mathscr{A}}
\nc{\B}{\mathscr{B}}
\nc{\C}{\mathscr{C}}
\nc{\D}{\mathscr{D}}
\nc{\Z}{\mathcal{Z}}
\nc{\ep}{\varepsilon}
\nc{\chem}{\mu_{\rm c}}
\rnc{\(}{\left(}
\rnc{\)}{\right)}
\rnc{\[}{\left[}
\rnc{\]}{\right]}
\nc{\dprime}{\prime\prime}
\nc{\bi}{{\overline{i}}}
\nc{\bT}{{\overline{T}}}
\nc{\bR}{{\overline{R}}}
\nc{\tmfb}{\widetilde{\mfb}}
\nc{\bmfb}{\widehat{\mfb}}
\nc{\hGamma}{\widehat{\Gamma}}
\nc{\tGamma}{\widetilde{\Gamma}}
\nc{\lam}{\lambda}
\nc{\bxi}{\overline{\xi}}
\nc{\txi}{\widetilde{\xi}}
\nc{\tlam}{\widetilde{\lambda}}
\nc{\hlam}{\widehat{\lambda}}
\nc{\hxi}{\widehat{\xi}}
\nc{\tC}{\widetilde{\mathcal{C}}}
\nc{\tM}{\widetilde{M}}
\nc{\hM}{\widehat{M}}
\nc{\bM}{\overline{M}}
\nc{\tY}{\widetilde{Y}}
\nc{\hY}{\widehat{Y}}
\nc{\bY}{\overline{Y}}
\nc{\ckxi}{\check{\xi}}

\nc{\XXZ}{$XXZ$\,}
\nc{\sotimes}{\mathop{\otimes}_{\rm s}}

\DeclareMathOperator{\sh}{sh}
\DeclareMathOperator{\ch}{ch}
\DeclareMathOperator{\Tr}{Tr}
\DeclareMathOperator{\Str}{Str}

\newtheorem{theorem}{Theorem}
\newtheorem{proposition}{Proposition}

\newtheorem{corollary}{Corollary}

\numberwithin{equation}{section}
\numberwithin{lemma}{section}
\numberwithin{proposition}{section}
\numberwithin{theorem}{section}
\numberwithin{corollary}{section}

\begin{document}%
%
\title{Correlation functions of an interacting spinless \\ 
fermion model at finite temperature}
\author{Kohei Motegi\thanks{E-mail: motegi@gokutan.c.u-tokyo.ac.jp} \,
and  Kazumitsu Sakai\thanks{E-mail: sakai@gokutan.c.u-tokyo.ac.jp}\\\\
\it Institute of physics, University of Tokyo, \\
\it Komaba 3-8-1,
Meguro-ku, Tokyo 153-8902, Japan \\\\
\\}

\date{\today}
 
 
\maketitle
%
%
\begin{abstract}
We formulate correlation functions for a one-dimensional
interacting spinless fermion model at finite temperature.
By combination of a lattice path integral formulation for 
thermodynamics with the algebraic Bethe ansatz for fermion 
systems, the equal-time one-particle Green's function at arbitrary
particle density is expressed as a multiple integral 
form.
Our formula reproduces previously known results
in the following three limits: the zero-temperature, 
the infinite-temperature and the free fermion limits.
\end{abstract}

%
\section{Introduction}
%
The exact computation of correlation functions for strongly correlated 
quantum systems has been one of the major problems for years. 
Although, in general,  this is exceedingly difficult to achieve, 
several analytical approaches especially in 1D quantum integrable 
models have been provided to derive exact or manageable expressions 
of correlation functions. 
For instance, the low-energy behavior of  correlation functions
for gapless models can be systematically obtained by conformal field 
theory \cite{BPZ,KY,FK,KBIbook}.  On the other hand, for systems with 
finite spectral gaps, the long-distance and -time asymptotics are 
investigated by (finite-temperature) form factor expansions 
(see \cite{D,AKT} for recent developments).

An alternative approach, which has been developed in these several years 
particularly for the  spin-1/2 XXZ chain,  is to combine the algebraic 
Bethe ansatz  \cite{KBIbook} with  solutions to the quantum inverse 
problem for local spin operators \cite{KMT99}. 
Using this, Kitanine {\it et al} derived multiple integral
representations for zero-temperature correlation functions of the 
XXZ chain with an external field \cite{KMT00,KMST02,KMSTreview}. 
Their representations can be regarded as natural extensions of the 
results based on the $q$-vertex operator approach \cite{JMMN,JMbook,JM},
which is restricted  to the zero-magnetic field case.
One of the  advantages of this method is that the formulation
can be flexibly  generalized to finite temperature and/or time 
dependent  case \cite{KMST05,GKS04,GHS05,Sakai07} by combining a lattice
path integral formulation. Furthermore, by considering
a continuum limit of the XXZ chain, correlation functions 
of the 1D boson system with delta function interaction can be
obtained at finite temperature \cite{SBGK07} 
(see also \cite{KKMST07} for the zero-temperature case).

Beyond spin systems,  more recently we further extended the method 
to the calculation of correlation functions for fermion systems.
By use of the fermionic $R$-operator \cite{USW} acting directly on
the fermionic Fock space, we have derived  multiple integral 
representations of zero-temperature correlation 
functions for an interacting  spinless fermion model with arbitrary 
particle density  \cite{MS}. 
In this paper, we generalize the former results to the finite-temperature
case by use of the quantum transfer matrix technique utilizing a concept of 
path integral \cite{SSSU}.  Especially considered here is the 
equal-time one-particle Green's function.
Our formula  agrees with  previously known results in the following 
three limits: the zero-temperature, infinite-temperature and the free 
fermion limits.

The layout of the paper is as follows. In the next section, we review 
the quantum transfer matrix method for the spinless fermion model,
and express the correlation function in terms of matrix elements
of the monodromy operator. In section 3, we present the key ingredients 
of the computation for the correlation function.   The multiple 
integral representation for the equal-time one-particle Green's
function at finite temperature with arbitrary particle density
is summarized in the main theorem.
In section 4, the three special limits are evaluated.
Section 5 is devoted to a brief discussion. The detailed 
derivation of the multiple integral form is deferred to  
the appendix.
%
\section{Spinless fermion model}
%
In this section, the thermodynamics of an interacting 
spinless fermion model is formulated by the quantum 
transfer matrix method \cite{SSSU}. The two-point correlation 
functions at finite temperature are expressed in terms of matrix 
elements of the monodromy operator.
%
\subsection{Fermionic $R$-operator}

The Hamiltonian of the interacting spinless fermion model on a 1D periodic 
lattice with $L$ sites is defined as
\begin{align}
&H=H_{0}-\mu_{\rm c} \sum_{j=1}^{L}\(\frac{1}{2}-n_{j}\), \nonumber \\
&H_{0}=t\sum_{j=1}^{L}\left\{ c_{j}^{\dagger}c_{j+1}+c_{j+1}^{\dagger}
                              c_{j}+2\Delta\(\(\frac{1}{2}-n_{j}\)
                                  \(\frac{1}{2}-n_{j+1}\)-\frac{1}{4}\)
                     \right\},
\label{hamiltonian}
\end{align}
where $c_{j}^{\dagger}$ and $c_{j}$ are the fermionic creation and 
annihilation operators at the $j$th site, respectively, satisfying the 
canonical anti-commutation relations.
Here $t$ and $\Delta$ are real  constants characterizing
the nature of the ground state, and $\chem$ denotes the
chemical potential coupling to the density operator 
$n_j=c_j^{\dagger}c_j$.

The underlying integrability of the model \eqref{hamiltonian}
can be seen by introducing  the fermionic $R$-operator defined as
\begin{equation}
R_{\bi j}(\lambda)=1-n_{\bi }-n_{j}+\frac{\sh\lambda}{\sh(\lambda+\eta)}
 (n_{\bi }+n_j-2 n_{\bi } n_j)
 +\frac{\sh\eta}{\sh(\lambda+\eta)}
 (c_{\bi }^{\dagger}c_{j}+c_{j}^{\dagger}c_{\bi }),
\label{fermion-R1}
\end{equation}
which acts on $V_{\bi } \sotimes V_{j}$. Here $V_{k}$ is a 
two-dimensional fermion Fock space whose normalized orthogonal
basis is given by
$|0 \ket _{k}$ and $|1 \ket _{k} := c_{k}^{\dagger} | 0\ket_{k}$,
where $ c_{k}|0 \ket_{k}=0$  and  $\sotimes$ denotes the super tensor product.
Identifying the above Fock space $V_j$ (respectively 
$V_{\bi }$) with the quantum space $\mathcal{H}_j$ (respectively 
the auxiliary space $\mathcal{H}_{\bi }$), we define the monodromy 
operator $\mathcal{T}_{\bi }^{\rm R}(\lambda)$ acting on the
space $V_{\bi } \sotimes (V_1\sotimes V_2 \sotimes \dots 
\sotimes V_L)$ as
\begin{equation}
\mathcal{T}_{\bi }^{\rm R}(\lambda)= 
R_{\bi L}(\lambda)\dots R_{\bi 2}(\lambda) R_{\bi 1}(\lambda). \nn
\end{equation}
Since the fermionic $R$-operator \eqref{fermion-R1} satisfies 
the Yang-Baxter equation
\cite{USW}
\begin{equation}
R_{12}(\lambda_{1}-\lambda_{2})R_{13}(\lambda_{1})R_{23}(\lambda_{2})
=R_{23}(\lambda_{2})R_{13}(\lambda_{1})R_{12}(\lambda_{1}-\lambda_{2}), \nn
\end{equation}
the transfer operator 
\begin{equation}
T_{\rm R}(\lambda)=\Str_{\bi } \mathcal{T}^{\rm R}_{\bi }(\lambda)
={}_{\bi } \bra 0|\mathcal{T}^{\rm R}(\lambda) |0 \ket_{\bi } 
-{}_{\bi }\bra 1|\mathcal{T}^{\rm R}(\lambda) |1 \ket_{\bi} \nn
\end{equation}
constitutes a commuting family:
[$T_{\rm R}(\lambda), T_{\rm R}(\mu)]=0$,
where the dual fermion Fock space is spanned 
by $_{k}\bra 0|$ and $_{k}\bra 1|$ with
${}_{k}\bra 1|:= {}_{k} \bra 0|c_{k}$
and
${}_{k}\bra 0|c_{k}^{\dagger}=0$.
The Hamiltonian $H_0$ \eqref{hamiltonian} is expressed in 
terms of the logarithmic derivative of the transfer operator 
$T_{\rm R}(\lambda)$:
\begin{equation}
H_{0}=t \sh (\eta) \frac{\partial}{\partial \lambda}
        \ln T_{\rm R}(\lambda)\biggr|_{\lambda=0},
\quad \Delta=\ch \eta.                       \nn
\end{equation}
This relation yields
\begin{equation}
T_{\rm R}(\lambda)=T_{\rm R}(0)
         \(1+\frac{\lambda}{ t\sh \eta} H_0 +\mathcal{O}(\lambda^2)\).
\label{expansion}
\end{equation}

For later use, let us define another type of transfer operator $\bT_{\rm R}(\lambda)$ 
\cite{SSSU}:
\begin{equation}
\bT_{\rm R}(\lambda)=
  \Str_{\bi}[\bR_{\bi L}(-\lambda)\dots\bR_{\bi 2}(-\lambda)\bR_{\bi 1}(-\lambda)] \nn
\end{equation}
with
\begin{equation}
\bR_{\bi j}(\lambda)=R_{j \bi}^{{\rm st}_j}(\lambda)=
  1-n_{\bi }-n_{j}+\frac{\sh\lambda}{\sh(\lambda+\eta)}
      (n_{\bi }+n_j-2 n_{\bi } n_j)
 -\frac{\sh\eta}{\sh(\lambda+\eta)}
 (c_{\bi }^{\dagger}c_{j}^{\dagger}-c_{j}c_{\bi }),
\label{fermion-R2}
\end{equation}
where ${\rm st}_j$ denotes the supertranspose with 
respect to the $j$th space.
Note that $T_{\rm R}(0)$ ($\bT_{\rm R}(0)$)
is the right-shift (left-shift) operator, namely
$T_{\rm R}(0) x_j=x_{j+1}T_{\rm R}(0)$ 
($\bT_{\rm R}(0)x_j=x_{j-1}\bT_{\rm R}(0)$)
where $x_j=c_j, c_j^{\dagger}$, and hence
$T_{\rm R}^{-1}(0)=\bT_{\rm R}(0)$.
Using this together with the expansion \eqref{expansion}, one
finds the statistical operator $\e^{-H/T}$ ($T$: temperature)
is given by
\begin{equation}
\e^{-H/T}=   \e^{\sum_{j=1}^L \mu_{\rm c}(1-2n_j)/(2T)}
 \lim_{N\to\infty}
   \left[\bT_{\rm R}(\lambda)
       T_{\rm R}\(\lambda-\frac{\beta}{N}\)\right]^{\frac{N}{2}}
       \Biggr|_{\lambda=0},
\qquad \beta=\frac{2 t \sh \eta}{T},
\label{boltzmann}
\end{equation}
where the Trotter number $N$ is assumed to be $N\in 2 \mathbb{N}$.
Note here that we have set the Boltzmann constant to unity.
%
%
\subsection{Correlation functions at finite temperature}
%
To derive multiple integral representations of finite
temperature correlation functions, here we  describe how 
the correlation functions can be expressed in terms
of the transfer operator formalism.

Let us consider a two-point 
correlation function at finite temperature $T>0$:
\begin{equation}
\bra \mathcal{O}_{m+1} \mathcal{O}_1^{\dagger} \ket
=\frac{\Tr \{ \e^{-H/T} \mathcal{O}_{m+1}
                     \mathcal{O}^{\dagger}_1
           \}
      }{\Tr \e^{-H/T}} \quad (m\ge 1),
\label{correlation}
\end{equation}
where $\mathcal{O}_j$ is a local fermion operator.
Inserting the formula \eqref{boltzmann} into the above,
and using the fact that the
$R$-operator \eqref{fermion-R1} or \eqref{fermion-R2} 
is Grassmann even, one obtains 
\begin{equation}
\bra \mathcal{O}_{m+1}\mathcal{O}^{\dagger}_1\ket
=\lim_{N\to\infty}\frac{\Str_{\overline{1},\dots,\overline{N}}\Tr_{1,\dots,L}
      \{ \mathcal{T}_L(0) \cdots 
       (\mathcal{T}_{m+1}(0)\mathcal{O}_{m+1})
       \mathcal{T}_m(0) \cdots 
       (\mathcal{T}_1(0) \mathcal{O}^{\dagger}_1)\}}
      {\Str_{\overline{1},\dots,\overline{N}}\Tr_{1,\dots,L}
       \{\mathcal{T}_L(0) \cdots 
       \mathcal{T}_1(0)\}}, \nn
\end{equation}
where the operator $\mathcal{T}_{j}(\lambda)$ acting
in the space 
$(V_{\overline{1}}\sotimes \cdots \sotimes V_{\overline{N}})\sotimes V_{j}$
is defined by
\begin{align}
\mathcal{T}_{j}(\lambda)&=
 \e^{\mu_{\rm c}(1-2n_j)/(2T)}
 \bR_{\overline{N}j}(-\lambda)R_{\overline{N-1}j}\(\lambda-\frac{\beta}{N}\)
\cdots
  \bR_{\overline{2}j}(-\lambda)R_{\overline{1}j}\(\lambda-\frac{\beta}{N}\) \nn \\
& = A(\lambda)(1-n_j)+B(\lambda)c_j+c_j^{\dagger} C(\lambda)+D(\lambda)n_j.  \nn
\end{align}
The Yang-Baxter equation and its modification
\begin{equation}
\bR_{31}(-\lambda_2)\bR_{32}(-\lambda_1)R_{12}(\lambda_1-\lambda_2)
=R_{12}(\lambda_1-\lambda_2)\bR_{32}(-\lambda_1)\bR_{31}(-\lambda_2) \nn
\end{equation}
yield
\begin{equation}
\mathcal{T}_1(\lambda_2)\mathcal{T}_2(\lambda_1)R_{12}(\lambda_1-\lambda_2)
=R_{12}(\lambda_1-\lambda_2)\mathcal{T}_2(\lambda_1)\mathcal{T}_1(\lambda_2), \nn
\end{equation}
and therefore the quantum transfer matrix defined by
\begin{equation}
T(\lambda)=\Tr_j \mathcal{T}_j(\lambda)=A(\lambda)+D(\lambda)
\label{qtm}
\end{equation}
commutes for different spectral parameters: $[T(\lambda),T(\mu)]=0$.
Thus \eqref{correlation} reduces to
\begin{equation}
\bra \mathcal{O}_{m+1}\mathcal{O}^{\dagger}_1\ket
=\lim_{N\to\infty}\frac{\Str_{\overline{1},\dots,\overline{N}}
      \{ T^{L-m-1}(0)\Tr_{m+1}\{\mathcal{T}_{m+1}(0)\mathcal{O}_{m+1}\}
         T^{m-1}(0) \Tr_{1}\{\mathcal{T}_1(0) \mathcal{O}_1^{\dagger}\}\}}
      {\Str_{\overline{1},\dots,\overline{N}}T^L(0)}.  \nn
\end{equation}

Let us consider the thermodynamic limit $L\to \infty$. Since the
two limits $L\to\infty$ and $N\to\infty$ are interchangeable
\cite{Suz,SuzIn}, we can take the limit $L\to\infty$ first.
In addition,  we find that the leading eigenvalue of the quantum
transfer matrix $T(0)$ (written as $\Lambda_0(0)$) is non-degenerate
and separated from the next-leading eigenvalues by a finite gap even
in the Trotter limit $N\to\infty$. 
In the thermodynamic limit $L\to\infty$, therefore, \eqref{correlation}
can be written in terms of $\Lambda_0(0)$ and the corresponding (normalized)
eigenstate $|\Psi_0\ket$ (note that 
$\Lambda_0(\lambda):=\bra \Psi_0|T(\lambda)|\Psi_0 \ket$). Namely
\begin{equation}
\bra \mathcal{O}_{m+1} \mathcal{O}_{1}^{\dagger}\ket
 =\lim_{N\to\infty}
  \frac{
   \bra \Psi_0| \Tr_{m+1}\{\mathcal{T}_{m+1}(0)\mathcal{O}_{m+1}\}
              (A+D)^{m-1}(0)
              \Tr_{1}\{\mathcal{T}_1(0)\mathcal{O}_1^{\dagger} \} 
   |\Psi_0 \ket
        }{\Lambda_0^{m+1}(0)}.
\label{correlation2}
\end{equation}
In particular, for the equal-time one-particle Green's function
(set $\mathcal{O}_j=c_j$), which will  mainly be 
considered in this paper, one obtains 
\begin{equation}
\bra c_{m+1} c_1^{\dagger}\ket=\bra c_1 c_{m+1}^{\dagger}\ket
=-\lim_{N\to\infty}\Lambda_0^{-m-1}(0) 
      \bra \Psi_0| C(0)(A+D)^{m-1}(0) B(0)  
   |\Psi_0 \ket.
\label{green}
\end{equation}
Note that  we have used $\Tr_j\{\mathcal{T}_j(0)c_j\}=-C(0)$
and $\Tr_j\{\mathcal{T}_j(0)c_j^{\dagger}\}=B(0)$.
%
\subsection{Diagonalization of the quantum transfer matrix}
%
To evaluate the correlation function \eqref{correlation2} (or \eqref{green})
actually, one must investigate the leading eigenvalues $\Lambda_0(0)$ 
and the corresponding eigenstates $|\Psi_0\ket$ 
of the quantum transfer matrix $T(0)$.
In this subsection, we present a general formula describing 
the eigenvalues and the eigenstates of $T(\lambda)$. The leading
eigenvalue $\Lambda_0(0)$ is expressed via the solution to
a nonlinear integral equation.

Let us define the reference state $|\Omega\ket$ as
\begin{equation}
|\Omega\ket:=|0\ket_{\overline{1}}\sotimes |1\ket_{\overline{2}} \sotimes
             \cdots  |0\ket_{\overline{N-1}}\sotimes |1\ket_{\overline{N}}.
\label{vacuum}
\end{equation}
Obviously \eqref{vacuum} is an eigenstate of $T(\lambda)$ \eqref{qtm}:
\begin{equation}
T(\lambda)=(a(\lambda)+d(\lambda))|\Omega\ket,
\quad A(\lambda)|\Omega\ket=a(\lambda)|\Omega\ket, \quad
      D(\lambda)|\Omega\ket=d(\lambda) |\Omega\ket,    \nn
\end{equation}
where
\begin{equation}
a(\lambda)=\left\{ \frac{\sh\lambda}{\sh(\lambda-\eta)}\right\}^{\frac{N}{2}}
                        \e^{\frac{\mu_{\rm c}}{2 T}},
   \quad 
d(\lambda)=(-1)^{\frac{N}{2}}\left\{ \frac{\sh(\lambda-\frac{\beta}{N})}
                        {\sh(\lambda-\frac{\beta}{N}+\eta)}\right\}^{\frac{N}{2}}
                        \e^{-\frac{\mu_{\rm c}}{2 T}}.     \nn
\end{equation}
In the framework of the algebraic Bethe ansatz, the
vector $|\{\lambda\}\ket$  constructed by the multiple 
action of $B(\lambda)$ on $|\Omega\ket$, namely
$
|\{\lambda\}\ket=\prod_{j=1}^M B(\lambda_j)| \Omega \ket
$,
is an eigenstate of $T(\lambda)$ if the complex parameters
$\{\lambda_j \}_{j=1}^M$
satisfy the  Bethe ansatz equation:
\begin{equation}
\frac{a(\lambda_j)}{d(\lambda_j)}=-(-1)^M \prod_{k=1}^M 
                      \frac{\sh(\lambda_j-\lambda_k+\eta)}
                           {\sh(\lambda_j-\lambda_k-\eta)}.
\label{bae}
\end{equation}
The corresponding eigenvalue of $T(\lambda)$ is written as
\begin{equation}
\Lambda(\lambda)=
a(\lambda)\prod_{j=1}^M\frac{\sh(\lambda-\lambda_j-\eta)}
                                                {\sh(\lambda-\lambda_j)}+
                    (-1)^M d(\lambda)\prod_{j=1}^M\frac{\sh(\lambda-\lambda_j+\eta)}
                                                {\sh(\lambda-\lambda_j)}.
\label{eigenvalue}
\end{equation}

\begin{figure}[ttt]
\begin{center}
\includegraphics[width=0.46\textwidth]{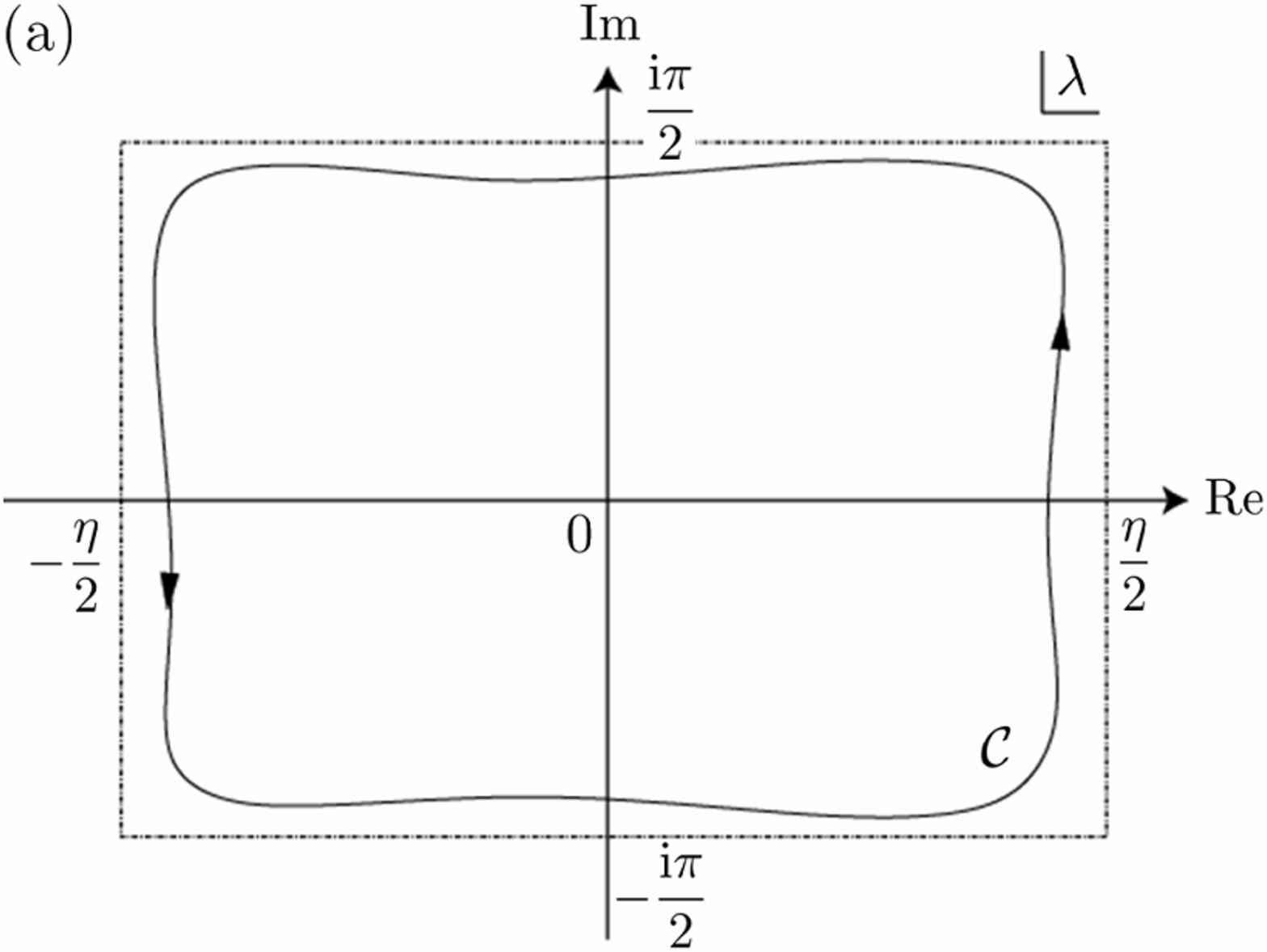}
\includegraphics[width=0.46\textwidth]{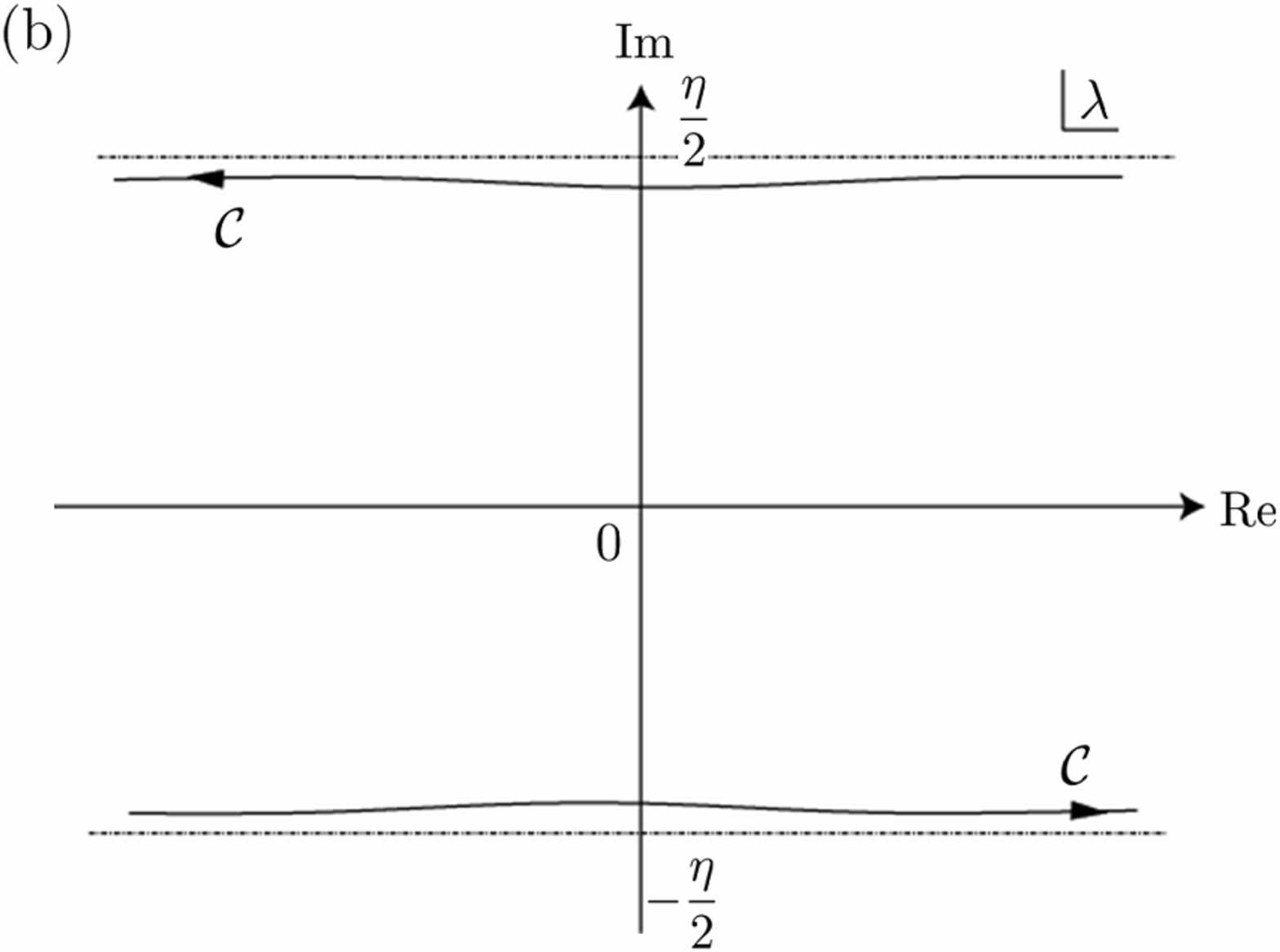}
\end{center}
\caption{The integration contours for the off-critical
regime $\Delta>1$ (a) and for the critical regime
$0\le \Delta \le 1$ (b). }
\label{contour-pic}
\end{figure}

The Bethe roots $\{\lambda \}$ characterizing the leading eigenvalue 
$\Lambda_0(0)$ are given by solutions to the Bethe ansatz 
equation \eqref{bae} in the sector $M=N/2$. 
Then the Bethe ansatz equation \eqref{bae} and the eigenvalue formula 
\eqref{eigenvalue} are exactly the same as the spin-1/2 XXZ chain
given by the Jordan-Wigner transformation, and hence
we can directly utilize the method  as in 
\cite{Klumper92,Klumper93,Destri92}, which makes the analysis
possible even in the Trotter limit $N\to\infty$.
Let us consider the following auxiliary function
\begin{equation}
\mfa(\lambda)=(-1)^{\frac{N}{2}}
\frac{d(\lambda)}{a(\lambda)}\prod_{k=1}^{N/2} 
                      \frac{\sh(\lambda-\lambda_k+\eta)}
                           {\sh(\lambda-\lambda_k-\eta)},
\label{auxiliary}
\end{equation}
which associates the Bethe roots $\{\lambda_j\}_{j=1}^{N/2}$
with zeros of $1+\mfa(\lambda)$. To study the analytical properties 
of this function, we need to know the distribution of the
Bethe roots describing the leading eigenvalue.
It has been numerically verified for a wide range of Trotter numbers
that the roots are distributed inside $\mathcal{C}$. For instance,
at $\mu_{\rm c}=0$, the Bethe roots for the critical (off-critical)
regime are located on the real (imaginary) axis.
Thus we can safely assume the 
above features hold for any Trotter numbers,
from which the analytical properties of the auxiliary function
are determined.
Consequently one sees $\mfa(\lambda)$
satisfies the following nonlinear
integral equation:
\begin{align}
\ln\mfa(\lambda)=&-\frac{\mu_{\rm c}}{T}+
\frac{N}{2}
\ln\frac{\sh(\lambda+\eta)\sh(\lambda-\frac{\beta}{N})}
        {\sh(\lambda)\sh(\lambda-\frac{\beta}{N}+\eta)}
-\int_{\mathcal{C}}\frac{\d\omega}{2\pi\i}\frac{\sh(2\eta)\ln(1+\mfa(\omega))}
             {\sh(\lambda-\omega+\eta)\sh(\lambda-\omega-\eta)}. 
\label{nlie-f}
\end{align} 
Here the contour $\mathcal{C}$ is taken, for instance, as a
rectangular contour whose edges are parallel to the
real axis at $\pm \pi \i/2$ 
(respectively $\pm \eta/2$) and are parallel
to the imaginary axis at $\pm \eta/2$ (respectively
$\pm \infty$) for
the off-critical regime $\Delta=\ch\eta>1$ (respectively for
the critical regime $0\le \Delta=\ch\eta\le 1$)
(see figure~\ref{contour-pic} for a pictorial definition).
In \eqref{nlie-f}
the Trotter limit $N\to\infty$ 
can be taken analytically:
\begin{equation}
\ln\mfa(\lambda)=-\frac{\mu_{\rm c}}{T}-\frac{ t \sh^2(\eta)}
{T\sh(\lambda)\sh(\lambda+\eta)}-
\int_{\mathcal{C}}\frac{\d\omega}{2\pi\i}\frac{\sh(2\eta) 
                      \ln(1+\mfa(\omega))}
             {\sh(\lambda-\omega+\eta)\sh(\lambda-\omega-\eta)}.
\label{nlie1}
\end{equation}
For later convenience,
we also introduce another auxiliary function $\mfab(\lambda)
=1/\mfa(\lambda)$ satisfying the following 
nonlinear integral equation in the limits $N \to\infty$
\cite{GKS04}:
\begin{equation}
\ln\mfab(\lambda)=\frac{\mu_{\rm c}}{T}-\frac{ t \sh^2(\eta)}
{T\sh(\lambda)\sh(\lambda-\eta)}+
\int_{\mathcal{C}}\frac{\d\omega}{2\pi\i}\frac{\sh(2\eta) 
                      \ln(1+\mfab(\omega))}
             {\sh(\lambda-\omega+\eta)\sh(\lambda-\omega-\eta)}.
\label{nlie2}
\end{equation}

By the above auxiliary function $\mfa(\lambda)$, 
the leading eigenvalue $\Lambda_0(0)$ of the quantum transfer matrix
$T(0)$,
which is related to the free energy density $f$
by $f=-T\ln\Lambda_0(0)$, is expressed as the following
single integral form:
\begin{equation}
\ln\Lambda_0(0)=\frac{\mu_{\rm c}}{2T}+\int_{\mathcal{C}}
      \frac{\d\omega}{2\pi\i}
        \frac{\sh(\eta)\ln(1+\mfa(\omega))}
             {\sh(\omega)\sh(\omega+\eta)}
=-\frac{\mu_{\rm c}}{2T}-\int_{\mathcal{C}}
      \frac{\d\omega}{2\pi\i}
        \frac{\sh(\eta)\ln(1+\mfab(\omega))}
             {\sh(\omega)\sh(\omega-\eta)}.
\label{largest}
\end{equation}
Differentiating \eqref{largest} with respect to the chemical
potential $\mu_{\rm c}$, 
one obtains the particle density $\bra n_j \ket$
\begin{equation}
\bra n_j \ket=-\int_{\mathcal{C}}
      \frac{\d\omega}{2\pi\i}
        \frac{T\sh(\eta)  \partial_{\mu_{\rm c}} \mfa(\omega)}
             {\sh(\omega)\sh(\omega+\eta)(1+\mfa(\omega))}
 =1+\int_{\mathcal{C}}
      \frac{\d\omega}{2\pi\i}
        \frac{T\sh(\eta) \partial_{\mu_{\rm c}} \mfab(\omega)}
             {\sh(\omega)\sh(\omega-\eta)(1+\mfab(\omega))}.  \nn
\end{equation}
%
%
%
%
\section{Multiple integral representation}
%
Along the line developed in \cite{GKS04},  we can  derive a 
multiple integral representing the  equal-time one-particle 
Green's function for the spinless fermion model. 
Here we sketch briefly how to derive  the multiple
integral by presenting some crucial formulae to evaluate 
the action of the operator $A+D$ and the resultant
scalar product. These formulae are essentially the 
same with those for the zero-temperature case \cite{MS}, 
since the commutation relations of the operators
$A$, $B$, $C$ and $D$ are exactly the same with
those for the zero-temperature case.
First, it is convenient to introduce the following more general 
function $\Phi_N(\{\xi\})$ instead of \eqref{green}:
\begin{align}
\Phi_N(\{\xi\})&=-\frac{\bra \Psi_0| C(\xi_1) \prod_{j=2}^{m}(A+D)(\xi_j)
                                  B(\xi_{m+1})
                       |\Psi_0 \ket}{\prod_{j=1}^{m+1}\Lambda_0(\xi_j)}  \nn \\
               &=-\frac{\bra \{\lam\}| C(\xi_1) \prod_{j=2}^{m}(A+D)(\xi_j)
                                  B(\xi_{m+1}) 
                       |\{\lam\}\ket}
                     {\bra \{\lam\}|\{\lam\} \ket\prod_{j=1}^{m+1}\Lambda_0(\xi_j)},
\label{inhom}                     
\end{align}
where $\{\xi_j\}_{j=1}^{m+1}$ is complex parameters located
inside $\mathcal{C}$. Note that $\{\lam\}$ and $|\{\lam\}\ket$ are, 
respectively, the Bethe roots
and the eigenvector (not normalized), which  characterize the 
leading  eigenvalue $\Lambda_0(0)$ (see the preceding section).
The dual vector $\bra \{\lam \}|$ is constructed by the multiple action of 
$C(\lam)$ on the state $\bra \Omega|$: 
$\bra \{\lam\}|=\bra \Omega|\prod_{j=1}^{N/2}C(\lam_j)$.
It immediately follows that the one-particle Green's function 
\eqref{green} can be obtained by taking the homogeneous limit 
$ \{ \xi \} \to 0$ and  the Trotter limit $N \to \infty $ 
in \eqref{inhom}:
\begin{equation}
\bra c_1 c_{m+1}^{\dagger} \ket=\lim_{N\to\infty}\lim_{\xi \to 0}
\Phi_{N}(\{\xi\}).
\label{green2}
\end{equation}

To evaluate the multiple action of the operator $A+D$ on the state 
$\bra \{\lambda\}| C(\xi_1)$,  let us introduce the
following proposition, which is originally proposed in
the calculation of the correlation function for the spin-1/2
XXZ chain \cite{KMST02}.
\begin{proposition} {\rm \cite{MS}}.
The action of 
$\prod_{j=1}^{m}(A+\kappa D)(\xi_{j}) $ 
on a state 
$\bra \Omega | \prod_{j=1}^{M}C(\mu_{j})=
\bra \{ \mu \} |$,
for any sets of complex parameters $\{\mu_j\}_{j=1}^M$
(not necessarily the Bethe roots),
is written as
\begin{align}
\bra \{  \mu \}|\prod_{j=1}^m &(A+\kappa D)(\xi_j) =
\sum_{n=0}^{p} \sum_{\substack{\{\mu \}=\{\mu^+\}\cup\{\mu^-\}\\
                             \{\xi\}=\{\xi^+\}\cup\{\xi^-\}\\
                             |\mu^+|=|\xi^+|=n}}
          R_n(\{\xi^+\}|\{\xi^-\}|\{\mu^+\}|\{\mu^-\})
        \bra \{\xi^{+} \} \cup \{  \mu^{-} \}|, \nn
\end{align}
where $p=\min(m,M)$, $
\bra \{\xi^{+} \} \cup \{  \mu^{-} \}|=\bra \Omega|\prod_{j=1}^{n} C(\xi^+_j)
                     \prod_{k=1}^{M-n}C(\mu_k^-)$ 
and the coefficient $R_n$ is given by
\begin{align}
R_n(\{\xi^+\}|\{\xi^-\}|\{\mu^+\}|\{\mu^-\})=&
     S_{n}(\{\xi^+\}|\{\mu^+\}|\{\mu^-\}) 
       \prod_{j=1}^{m-n} \biggl[ a(\xi^-_{j})\prod_{k=1}^{n}f(\xi^+_{k},\xi^-_{j})
       \prod_{k=1}^{M-n}f(\mu^-_{k},\xi^-_{j}) \nn \\
&
 +\kappa d(\xi^-_{j})\prod_{k=1}^{n}\{-f(\xi^-_{j},\xi^+_{k}) \} 
                    \prod_{k=1}^{M-n} \{-f(\xi^-_{j},\mu^-_{k}) \} \biggr].
\label{R}
\end{align}
Here $S_{n}$ is defined as
\begin{equation}
S_{n}( \{ \xi^+ \}| \{ \mu^+ \}| \{ \mu^- \})=
                  \frac{\prod_{j,k=1}^{n} 
                        \sh  (\xi^+_j-\mu^+_k+\eta)}
                   {\prod_{j<k}^{n}\left[\sh(\mu^+_k-\mu^+_j)
                    \sh(\xi^+_j-\xi^+_k)\right]}{\det}_n M_{jk}  \nn
\end{equation}
with
\begin{align}
M_{jk}=&a(\mu^+_{j})t(\xi^+_{k},\mu^+_{j})\prod_{a=1}^{M-n}
                    f(\mu^-_{a},\mu^+_{j})              \nn \\
&-\kappa d(\mu^+_{j})t(\mu^+_{j},\xi^+_{k})
       \prod_{a=1}^{M-n}\{-f(\mu^+_{j},\mu^-_{a}) \}
       \prod_{b=1}^{n} \left\{-\frac{\sh(\mu^+_{j}-\xi^+_{b}+\eta)}
                           {\sh  (\mu^+_{j}-\xi^+_{b}-\eta)} \right\}.
\label{M}
\end{align}
The functions $f(\lam,\mu)$ and $t(\lam,\mu)$ appearing in \eqref{R} and
\eqref{M} are, respectively, given by
\begin{equation}
f(\lam,\mu)=\frac{\sh(\lam-\mu+\eta)}{\sh(\lam-\mu)},
    \quad t(\lam,\mu)=\frac{\sh\eta}{\sh(\lam-\mu)\sh(\lam-\mu+\eta)}.  \nn
\end{equation}
\end{proposition}

Compared with that for the XXZ chain \cite{KMST02}, some sign factors appear 
in the second term of \eqref{R} and \eqref{M}, which 
originate from the fermionic nature of the present system.
By setting $\kappa=1$ and applying the above formula to \eqref{inhom}, 
one has
\begin{align}
& \Phi_N(\xi)=
-\sum_{n=0}^{m-1} \sum_{\substack{\{\widetilde{\lambda}\}=
\{\widetilde{\lambda}^+\}\cup\{\widetilde{\lambda}^-\}\\
                             \{\widetilde{\xi}\}=\{\widetilde{\xi}^+\}
                             \cup\{\widetilde{\xi}^-\}\\
                             |\widetilde{\lambda}^+|=|\widetilde{\xi}^+|=n}}
        \frac{  R_n(\{\widetilde{\xi}^+\}|\{\widetilde{\xi}^-\}
        |\{\widetilde{\lambda}^+\}|\{\widetilde{\lambda}^-\})
        \bra  \{ \widetilde{\xi}^+  \}  \cup \{  \widetilde{\lambda}^-  \}  | 
         B(\xi_{m+1})|\{ \lambda \}  \ket }
 {\bra
 \{ \lambda \} | \{ \lambda \} 
 \ket   
 \prod_{j=1}^{m+1} \Lambda_{0}(\xi_{j})}.
\label{mult2}
\end{align}
Here some new notations are adopted:
\begin{equation}
(\tlam_1,\dots,\tlam_{N/2+1})=(\lambda_{1},\dots,\lambda_{N/2},\xi_{1}),
\quad
(\txi_1,\dots,\txi_{m-1})=( \xi_{2},\dots, \xi_{m}).   \nn
\end{equation}

Next we evaluate the action of $B(\xi_{m+1})$ on 
$\bra \{\txi^+\}\cup\{\tlam^-\}|$ by using the  formula 
\cite{MS}:
\begin{align}              
    \bra \Omega|\prod_{j=1}^{M}C(\mu_{j})B(\mu_{M+1})=&
         (-1)^{M-1}\sum_{l=1}^{M+1}d(\mu_{l})\frac{\prod_{k=1}^{M}
                 \sh(\mu_{l}-\mu_{k}+\eta)}
                {\prod_{\substack{k=1 \\  k \neq l}}^{M+1}
                 \sh(\mu_{l}-\mu_{k})} \nn \\
 \times \sum_{\substack{l'=1\\l'\neq l}}^{M+1}
     & \frac{a(\mu_{l'})}{\sh(\mu_{M+1}-\mu_{l'}+\eta)}
     \frac{\prod_{\substack{j=1 \\ j\neq l}}^{M+1}
                      \sh(\mu_{j}-\mu_{l^{\prime}}+\eta)}
           {\prod_{\substack{j=1 \\  j \neq l,l^{\prime}}}^{M+1}
                    \sh(\mu_{j}-\mu_{l^{\prime}})} 
\bra \Omega| \prod_{\substack{j=1 \\ j \neq l,l^{\prime}}}^{M+1}
                                          C(\mu_{j}),
\label{action}
\end{align}               
where $\{\mu_j\}_{j=1}^{M+1}$  are arbitrary complex numbers.
One sees  that the resulting equation consists of the ratio 
of scalar products such as $\bra \{\xi^+\}\cup\{\lam^-\}| \{\lam\} \ket/
\bra \{\lam\}|\{\lam\} \ket$, where $\{\lam\}=\{\lam^+_j\}_{j=1}^n
\cup\{\lam^-_j\}_{j=1}^{N/2-n}$ and $\{\xi^+_j\}_{j=1}^n \in 
\{\xi_j\}_{j=1}^{m+1}$ (see the appendix for detail).
In fact, this quantity can be calculated by the following determinant
representation of the scalar product. 
\begin{proposition} \label{scalar} {\rm \cite{MS}}.
The scalar product between a Bethe state and an arbitrary state
\begin{equation}
\mathbb{S}_M(\{\mu\}|\{\lambda\})
=\bra \Omega| \prod_{j=1}^{M}C(\mu_{j})\prod_{j=1}^{M}
B(\lambda_{j})|\Omega \ket             \nn
\end{equation}
can be expressed as follows:
\begin{align}
\mathbb{S}_M(\{\mu\}|\{\lambda\}) 
  =(-1)^{\frac{M(M-1)}{2}}\frac{\prod_{j=1}^{M}d(\lambda_{j})a(\mu_{j}) 
 \prod_{j,k=1}^{M}\sh(\lambda_{j}-\mu_{k}+\eta)}
 {\prod_{j<k}^{M}\sh(\lambda_{j}-\lambda_{k})\sh(\mu_{k}-\mu_{j})}
  {\det}_{M}\Psi( \{ \mu \}| \{ \lambda \}),   \nn
\end{align}
where $\{ \lambda_{j} \}_{j=1}^M$  are Bethe roots,  $\{ \mu_{j} \}_{j=1}^M $ 
are arbitrary complex parameters.
The $M \times M$ matrix $\Psi(\{ \mu \} | \{ \lambda \} )$ 
is defined by       
\begin{equation}
\Psi_{jk}( \{ \mu \}| \{ \lambda \})=
  t(\lambda_{j},\mu_{k})-(-1)^{M}t(\mu_{k},\lambda_{j})\frac{d(\mu_{k})}{a(\mu_{k})}
  \prod_{a=1}^{M}   
      \frac{\sh(\mu_{k}-\lambda_{a}+\eta)}{\sh(\mu_{k}-\lambda_{a}-\eta)},  \nn
\end{equation}
and ${\det}_M$ denotes the determinant of an $M\times M$ matrix.
\end{proposition}

Applying this, and using the same technique proposed in \cite{GKS04},
one obtains the ratio of the scalar products 
$\bra \{\xi^+\}\cup\{\lam^-\}| \{\lam\} \ket/
\bra \{\lam\}|\{\lam\} \ket$:
\begin{align}
\frac{\bra \{\xi^+\}\cup\{\lam^-\}| \{\lam\} \ket}
     {\bra \{\lam\}|\{\lam\} \ket}
=& 
  \prod_{j=1}^{n}\[\frac{a(\xi_{j}^{+})(1+\mfa(\xi_{j}))}
    {a(\lambda_{j}^{+})\mfa^{\prime}(\lambda_{j}^{+})} \]
  \prod_{j=1}^{\frac{N}{2}-n}\prod_{k=1}^{n}  
    \[ \frac{f(\lambda_{j}^{-},\xi_{k}^{+})}{f(\lambda_{j}^{-},\lambda_{k}^{+})}\] \nn \\
&\times
   \prod_{j,k=1}^{n} \[\frac{\sh(\lambda_{j}^{+}-\xi_{k}^{+}+\eta)}
                           {\sh(\lambda_{j}^{+}-\lambda_{k}^{+}+\eta)}\]
    \prod_{j < k}^{n}\[ \frac{\sh(\lambda_{j}^{+}-\lambda_{k}^{+})}
           {\sh(\xi_{j}^{+}-\xi_{k}^{+})}\] {\det}_n G(\lambda_{j}^{+},\xi_{k}^{+}),
\label{ratio}
\end{align}
where the function $G(\lambda,\xi)$ satisfies the following linear integral equation
\begin{equation}
G(\lambda,\xi)=t(\xi,\lambda)+\int_{\mathcal{C}}
\frac{\d \omega}{2 \pi \i}\frac{\sh(2 \eta)}
{\sh(\lambda-\omega+\eta)\sh(\lambda-\omega-\eta)}
\frac{G(\omega,\xi)}{1+\mfa(\omega)},
\label{G-function}
\end{equation}
which can  also be written in terms of $\mfab(\lam)$ as
\begin{equation}
G(\lambda,\xi)=-t(\lambda,\xi)
-\int_{\mathcal{C}}\frac{\d \omega}{2 \pi \i}
       \frac{\sh (2\eta)}{\sh (\lambda-\omega+\eta) \sh (\lambda-\omega-\eta)}
       \frac{G(\omega,\xi)}{1+\mfab(\omega)}.
\label{G-function2}
\end{equation} 

Applying all the steps described above, we find that
\eqref{inhom} can be reduced to sums over the partitions of the sets
$\{\lam \}$ and $\{\xi\}$, and its summand consists
of determinants of matrices constructed by functions of 
$\{\lam\}$ and  $\{\xi\}$ (see \eqref{F1new} for example).
In fact, by using the technique as in \cite{GKS04}, 
these sums can be  transformed to multiple integrals 
on the canonical contour $\mathcal{C}$,
where the Trotter limit can be taken analytically.
The derivation is straightforward but has a lot of steps,
here we only write down the final result.
Namely, the function $\Phi_N(\{\xi\})$ \eqref{inhom} is
represented by the following multiple integral:
\begin{align}
\Phi_N&(\{\xi\})
    =\sum_{n=0}^{m-1}\frac{(-1)^{m}}{n!(n+1)!}
\int_{\Gamma^{n+1}}\prod_{j=1}^{n+1}\frac{\d \zeta_{j}}{2 \pi \i}
\frac{\sh(\zeta_{j}-\xi_{1}-\eta)}{\mfb_{-}(\zeta_{j})\sh(\zeta_{j}-\xi_{m+1})}
\nn \\
&\times \int_{\mathcal{C}^n}\prod_{j=1}^{n}
\frac{\d \omega_{j}}{2 \pi \i(1+\mfa(\omega_{j}))}
\frac{\mfb_{-}(\omega_{j})\sh(\omega_{j}-\xi_{m+1})}{\sh(\omega_{j}-\xi_{1}-\eta)}
\int_{\mathcal{C}}\frac{\d \omega_{n+1}}{2 \pi \i (1+\mfab(\omega_{n+1}))}
\int_{\mathcal{C}}\frac{\d \omega_{n+2}}{2 \pi \i(1+\mfa(\omega_{n+2}))}
\nn \\
&
\times \frac{\prod_{j=1}^{n+1} \[ \sh(\omega_{n+1}-\zeta_{j}+\eta)
                 \sh(\omega_{n+2}-\zeta_{j}-\eta) \]}
             {\prod_{j=1}^{n} \[ \sh(\omega_{n+1}-\omega_{j}+\eta)
                    \sh(\omega_{n+2}-\omega_{j}-\eta) \]}
        \frac{W^{-}_{n}( \{ \omega \} | \{ \zeta \})}
             {\sh(\omega_{n+1}-\omega_{n+2}+\eta)}   \nn \\
&
\times {\det}_{n+1}M^{-}_{jk}(\{\omega\}|\{\zeta\})
        {\det}_{n+2}\[ G(\omega_{j}, \zeta_{1}),\dots,
            G(\omega_{j}, \zeta_{n+1}),G(\omega_{j},\xi_{m+1}) \],
\label{mult-inhom1}
\end{align}
where 
\begin{align}
&\mfb_{\pm}(\omega)=\prod_{j=2}^{m}\frac{\sh(\omega-\xi_{j})}
                                      {\sh(\omega-\xi_{j}\pm\eta)}, \nn \\
&W^{\pm}_{n}( \{ \omega \} | \{ \zeta \})=
 \frac{\prod_{j=1}^{n}
       \prod_{k=1}^{n+1}\sh(\omega_{j}-\zeta_{k}\pm\eta)
                        \sh(\zeta_{k}-\omega_{j}\pm\eta)}
      {\prod_{j=1}^{n}\prod_{k=1}^{n}\sh(\omega_{j}-\omega_{k}\pm\eta)
\prod_{j=1}^{n+1}\prod_{k=1}^{n+1}\sh(\zeta_{j}-\zeta_{k}\pm\eta)},
\label{b-W}
\end{align}
and $M^{-}(\{\omega\}|\{\zeta\})$ is an $(n+1)\times (n+1)$ matrix 
whose matrix elements are given by
\begin{equation}
M^{-}_{jk}= \begin{cases}
           t(\omega_j,\zeta_k)+t(\zeta_k,\omega_j)
               \prod_{a=1}^{n}\frac{\sh(\omega_a-\omega_j-\eta)}
                                   {\sh(\omega_j-\omega_a-\eta)} 
               \prod_{b=1}^{n+1} \frac{\sh(\omega_j-\zeta_b-\eta)}
                                      {\sh(\zeta_b-\omega_j -\eta)}
                                   & \text{ for $j\le n$}             \\
        t(\xi_1,\zeta_k)
                                   & \text{ for $j= n+1$}    
        \end{cases}.   \nn
\end{equation}
$\mfa(\lambda)(=1/\mfab(\lam))$ and $G(\lambda, \zeta)$ satisfy 
the integral equations \eqref{nlie-f} and \eqref{G-function}, 
respectively.
$\mathcal{C}$ is the canonical contour defined as in figure~\ref{contour-pic},
and $\Gamma$ encircles  ${\xi}$ and does not contain any other singularities.

The one-particle Green's function can be obtained from the above expression by 
taking the limits $\{ \xi \} \to 0$ and  $N \to \infty $
(see \eqref{green2}). 
The latter means to take $\mfa(\lambda)$ as a function satisfying 
\eqref{nlie1}.
We thus arrive at
\begin{theorem}
The equal-time one-particle Green's function of the spinless fermion model
 at finite temperature has the following multiple integral representation, 
\begin{align}
\bra &c_{1}c_{m+1}^{\dagger} \ket =
  \sum_{n=0}^{m-1}\frac{(-1)^{m}}{n!(n+1)!}
     \int_{\Gamma^{n+1}}\prod_{j=1}^{n+1}\frac{\d \zeta_{j}}{2 \pi \i}
        \( \frac{\sh(\zeta_{j}-\eta)}{\sh(\zeta_{j})} \)^{m}    \nn \\
 &
    \times \int_{\mathcal{C}^n}\prod_{j=1}^{n}
       \frac{\d \omega_{j}}{2 \pi \i(1+\mfa(\omega_{j}))}
               \( \frac{ \sh (\omega_{j})}{ \sh (\omega_{j}-\eta)} \)^{m}
    \int_{\mathcal{C}}\frac{\d \omega_{n+1}}{2 \pi \i (1+\mfab(\omega_{n+1}))}
\int_{\mathcal{C}}\frac{\d \omega_{n+2}}{2 \pi \i (1+\mfa(\omega_{n+2}))}
\nn \\
&\times \frac{\prod_{j=1}^{n+1} \[ \sh(\omega_{n+1}-\zeta_{j}+\eta)
\sh(\omega_{n+2}-\zeta_{j}-\eta) \]}
{\prod_{j=1}^{n} \[ \sh(\omega_{n+1}-\omega_{j}+\eta)
\sh(\omega_{n+2}-\omega_{j}-\eta) \]}
\frac{W^{-}_{n}( \{ \omega \} | \{ \zeta \})}{\sh(\omega_{n+1}-\omega_{n+2}+\eta)} 
\nn \\
&\times {\det}_{n+1}M^{-}_{jk}(\{\omega\} | \{\zeta\})\Bigr|_{\xi_{1} \to 0}
  {\det}_{n+2}\[ G(\omega_{j}, \zeta_{1}),\dots,
G(\omega_{j}, \zeta_{n+1}),G(\omega_{j},0) \],
\label{mult-green1}
\end{align}
where $\mfa(\lambda)=1/(\mfab(\lam))$ and 
$G(\lambda, \zeta)$ satisfy the integral equation 
\eqref{nlie1} and \eqref{G-function}, respectively.
$\mathcal{C}$ is the canonical contour and $\Gamma$ surrounds the point ${0}$.
\end{theorem}

Using the identity 
\begin{equation}
\frac{1}{1+\mfa(\omega)}=1-\frac{1}{1+\mfab(\omega)},
\label{decompose}
\end{equation}
we can convert the above multiple integral representation
into another form.
Namely, inserting the  decomposition \eqref{decompose} into 
the part $\prod_{j=1}^n 1/(1+\mfa(\omega_j))$ of \eqref{mult-inhom1},
and then  performing the integrals over $\zeta_j$, we transform 
them to  sums over the partition of the set $\{\xi\}$.
Resumming the results in a similar way as in the appendix,
we have
\begin{align} 
\Phi_N&(\{\xi\})=\sum_{n=0}^{m}\frac{(-1)^{n}}{n!(n+1)!}
       \int_{\Gamma^{n+1}}\prod_{j=1}^{n+1}\frac{\d \zeta_{j}}{2 \pi \i}
          \frac{\sh(\zeta_{j}-\xi_{1}+\eta)}
               {\mfb_{+}(\zeta_{j})\sh(\zeta_{j}-\xi_{m+1})}        \nn \\
&\times \int_{\mathcal{C}^n}\prod_{j=1}^{n}
\frac{\d \omega_{j}}{2 \pi \i(1+\mfab(\omega_{j}))}
\frac{\mfb_{+}(\omega_{j})\sh(\omega_{j}-\xi_{m+1})}{\sh(\omega_{j}-\xi_{1}+\eta)}
\int_{\mathcal{C}}\frac{\d \omega_{n+1}}{2 \pi \i(1+\mfab(\omega_{n+1}))}
\int_{\mathcal{C}}\frac{\d \omega_{n+2}}{2 \pi \i(1+\mfa(\omega_{n+2}))}
\nn \\
&\times \frac{\prod_{j=1}^{n+1} \[ \sh(\omega_{n+1}-\zeta_{j}+\eta)
\sh(\omega_{n+2}-\zeta_{j}-\eta) \]}
{\prod_{j=1}^{n} \[ \sh(\omega_{n+1}-\omega_{j}+\eta)
\sh(\omega_{n+2}-\omega_{j}-\eta) \]}
\frac{W^{+}_{n}( \{ \omega \} | \{ \zeta \})}{\sh(\omega_{n+1}-\omega_{n+2}+\eta)} 
\nn \\
&\times {\det}_{n+1}M^{+}_{jk}(\{\omega\}|\{\zeta\})
     {\det}_{n+2}\[ G(\omega_{j}, \zeta_{1}),\dots,
G(\omega_{j}, \zeta_{n+1}),G(\omega_{j},\xi_{m+1}) \],  \nn
\end{align}
where $\mfb_+(\omega)$ and $W^{+}_{n}( \{ \omega \} | \{ \zeta \})$
are defined in \eqref{b-W}. 
$M^{+}(\{\omega\} | \{\zeta\}) $ is an $(n+1)\times (n+1)$ matrix 
whose matrix elements are given by
\begin{equation}
M^{+}_{jk}= 
 \begin{cases}
     t(\zeta_{k},\omega_{j})+t(\omega_{j},\zeta_{k})
        \prod_{a=1}^{n}\frac{\sh(\omega_{a}-\omega_{j}+\eta)}
                            {\sh(\omega_{j}-\omega _{a}+\eta)}
        \prod_{b=1}^{n+1}\frac{\sh(\omega_{j}-\zeta_{b}+\eta)}
                        {\sh(\zeta_{b}-\omega_{j}+\eta)} &\text{\quad for $j\le n$} \\
        t(\zeta_{k},\xi_{1}) &\text{\quad for $j=n+1$}
  \end{cases}.  \nn
\end{equation}
Taking the homogeneous and the Trotter limits, we have  another multiple
integral representing the one-particle Green's function.
\begin{corollary}
The equal-time one-particle Green's function of the spinless fermion 
model at finite temperature has another multiple integral representation:
\begin{align}
\bra &c_{1}c_{m+1}^{\dagger} \ket=
    \sum_{n=0}^{m-1}\frac{(-1)^{n}}{n!(n+1)!}
         \int_{\Gamma^{n+1}}\prod_{j=1}^{n+1}\frac{\d \zeta_{j}}{2 \pi \i} 
              \( \frac{\sh(\zeta_{j}+\eta)}{\sh(\zeta_{j})} \)^{m}      \nn \\
&
      \times   \int_{\mathcal{C}^n}\prod_{j=1}^{n}
\frac{\d \omega_{j}}{2 \pi \i(1+\mfab(\omega_{j}))}
\(\frac{\sh(\omega_{j})}{\sh(\omega_{j}+\eta)} \)^{m} 
\int_{\mathcal{C}}\frac{\d \omega_{n+1}}{2 \pi \i(1+\mfab(\omega_{n+1}))}
\int_{\mathcal{C}}\frac{\d \omega_{n+2}}{2 \pi \i(1+\mfa(\omega_{n+2}))}
\nn \\
&\times \frac{\prod_{j=1}^{n+1} \[ \sh(\omega_{n+1}-\zeta_{j}+\eta)
\sh(\omega_{n+2}-\zeta_{j}-\eta) \]}
{\prod_{j=1}^{n} \[ \sh(\omega_{n+1}-\omega_{j}+\eta)
\sh(\omega_{n+2}-\omega_{j}-\eta) \]}
\frac{W^{+}_{n}( \{ \omega \} | \{ \zeta \})}{\sh(\omega_{n+1}-\omega_{n+2}+\eta)} 
\nn \\
&\times {\det}_{n+1}M^{+}_{jk}(\{\omega\}|\{\zeta\})\Bigr|_{\xi_{1} \to 0} {\det}_{n+2}
\[ G(\omega_{j}, \zeta_{1}),\dots,
G(\omega_{j}, \zeta_{n+1}),G(\omega_{j},0) \],
\label{mult-green2}
\end{align}
where $\mfab(\lambda)=1/\mfa(\lam)$ and 
$G(\lambda, \zeta)$ satisfy the integral equations 
\eqref{nlie2} and \eqref{G-function2}, respectively.
\end{corollary}

%
\section{Special cases}
%
In this section, we evaluate the  three special cases 
of the multiple integral representation: the zero-temperature
and the infinite-temperature and the free fermion limits.
%
\subsection{Zero-temperature limit} 
%
First let us consider the zero-temperature limit. Here we 
restrict ourselves on the off-critical case $\Delta>0$, and
set $\eta<0$ as in \cite{MS}. The critical case $0\le \Delta\le 1$,
of course, can be treated by just changing the definition
of the integration contour $\mathcal{C}$ as in figure~\ref{contour-pic}.

Shifting the variables in  \eqref{mult-green2} by
$\omega_j\to\omega_{j}-\eta/2$ and
$\zeta_{j}\to\zeta_j-\eta/2$, we deal with the
integrals on the contour $\Gamma_{\eta/2}$ and $\mathcal{C}_0\cup
\mathcal{C}_{\eta}$, where $\Gamma_{\eta/2}$ encircles the
point $\eta/2$; $\mathcal{C}_0$ and $\mathcal{C}_{\eta/2}$
are defined as $\mathcal{C}_0=[-\pi\i/2,\pi \i/2]$ and 
$\mathcal{C}_\eta=[\eta+\pi \i/2,\eta-\pi \i/2]$, respectively.
A close inspection of the auxiliary functions $\mfa(\lam)$ 
and $\mfab(\lam)$ for $\mu_{\rm c}>0$ and $\eta<0$ at the
zero-temperature limit $T\to 0$ leads to
\begin{equation}
\frac{1}{1+\mfa(\lambda-\frac{\eta}{2})} \stackrel{T \to 0}{\longrightarrow} 
 \begin{cases}
   1 &\text{for $\lambda \in \overline{\mathcal{L}}$ } \\
   0 &\text{for $\lambda \in \mathcal{L}$ }
  \end{cases},   \quad
\frac{1}{1+\mfab(\lambda-\frac{\eta}{2})} \stackrel{T \to 0}{\longrightarrow} 
\begin{cases}
   0 &\text{for $\lambda \in \overline{\mathcal{L}}$ } \\
   1 &\text{for $\lambda \in \mathcal{L}$ }
  \end{cases}, 
\label{reduction}
\end{equation}
where $\mathcal{L}=[-q_{\mu_{\rm c}},q_{\mu_{\rm c}}]$ 
and $\overline{\mathcal{L}}=
   (\mathcal{C}_0\cup\mathcal{C}_\eta)\setminus\mathcal{L}$.
Note that the Fermi point $q_{\mu_{\rm c}}$  is
an imaginary number ($\Im q_{\mu_{\rm c}}>0$)
depending on the chemical potential $\mu_{\rm c}$.
Substituting this into \eqref{G-function2} and 
shifting the variables as above, one has
\begin{equation}
G\(\lambda-\frac{\eta}{2},\zeta-\frac{\eta}{2}\)=
   -t(\lambda,\zeta)+
     \int_{-\mathcal{L}}\frac{\d \omega}{2 \pi \i}
              \frac{\sh(2 \eta)G(\omega-\frac{\eta}{2},\zeta-\frac{\eta}{2})}
                   {\sh(\lambda-\omega+\eta)\sh(\lambda-\omega-\eta)}.  \nn
\end{equation}
Comparing this  with equation (2.24) in \cite{MS}, one can
identify $G(\lam,\zeta)$ with the density function $\rho(\lam,\zeta)$:
\begin{equation}
G\(\lambda-\frac{\eta}{2},\zeta-\frac{\eta}{2}\)=2\pi \i \rho(\lam,\zeta).
\label{density}
\end{equation}
Inserting both \eqref{density} and \eqref{reduction} into
\eqref{mult-green2}, we finally obtain 
\begin{align}
\lim_{T \to 0}&\bra c_{1}c_{m+1}^{\dagger} \ket
=
-\sum_{n=0}^{m-1}\frac{1}{n!(n+1)!}\int_{\Gamma_{\frac{\eta}{2}}}
         \prod_{j=1}^{n+1}\frac{\d \zeta_{j}}{2\pi \i} 
    \int_{-\mathcal{L}} \d^{n+1}\omega  
    \int_{\overline{\mathcal{L}}} \d \omega_{n+2}
\prod_{j=1}^{n+1} \( \frac{\sh(\zeta_{j}+\frac{\eta}{2})}
{\sh(\zeta_{j}-\frac{\eta}{2})} \)^{m}                    \nn \\
& 
  \times \prod_{j=1}^{n} \( 
          \frac{\sh(\omega_{j}-\frac{\eta}{2})}{\sh(\omega_{j}+\frac{\eta}{2})}
                         \)^{m}
          \frac{\prod_{j=1}^{n+1} \[ 
                  \sh(\omega_{n+1}-\zeta_{j}+\eta)\sh(\omega_{n+2}-\zeta_{j}-\eta)
                                 \] }
                {\prod_{j=1}^{n} \[ 
                        \sh(\omega_{n+1}-\omega_{j}+\eta)
                        \sh(\omega_{n+2}-\omega_{j}-\eta) 
                                 \] } 
          \frac{W_{n}^{+}( \{ \omega \} | \{ \zeta \})}
               {\sh(\omega_{n+1}-\omega_{n+2}+\eta)}\nn \\
&
\times
        {\det}_{n+1}M^{+}_{jk}(\{\omega\}|\{\zeta\})
                            \Bigr|_{\xi_{1} \to \frac{\eta}{2}} 
        {\det}_{n+2}\[ 
          \rho(\omega_{j}, \zeta_{1}),\dots,\rho(\omega_{j}, \zeta_{n+1}),
          \rho(\omega_{j},\eta/2) 
                     \].   \nn
\end{align}
The above representation  completely agrees with equation (4.18) in \cite{MS}.
%
\subsection{Infinite-temperature limit}
%
Next we would like to deal with the infinite-temperature
case $T=\infty$, where the function $\Phi_N(\{\xi\})$
\eqref{mult-inhom1} does not depend on the Trotter number 
$N$ (note that $\mfa(\lam)=1$ and $\mfab(\lam)=1$).
Therefore the integrals can be explicitly 
evaluated by just applying the residue theorem to the poles 
of integrand. The result reads
 \begin{align}
 \lim\limits_{T \to \infty} \bra c_{1}c_{m+1}^{\dagger} \ket=0, \nn
\end{align}
as one expected.
%
\subsection{Free fermion point}
%
Finally we investigate the representation \eqref{mult-green1} at
the free fermion point $\Delta=0$. 
Set $\eta =\pi \i/2$. Then the integral kernel in \eqref{nlie1} and 
\eqref{G-function} becomes zero. Hence the two functions
$\mfa(\omega)(=1/\mfab(\omega))$  and $G(\omega,\zeta)$ can be
explicitly written as
\begin{equation}
 \mfa(\omega)=\exp \left\{ -\frac{1}{T} \( \mu_{\rm c}+
             \frac{2 \i t}{\sh(2\omega)}
                                      \) \right\},                
\quad G(\omega, \zeta)=-\frac{2}{\sh(2(\omega-\zeta))}.  \nn
\end{equation} 
Since $M_{jk}=0$ for $j \le n$, one observes all the
terms $n\ge 1$ vanish. Applying the decomposition
$1/(1+\mfab(\omega_1))=1-1/(1+\mfa(\omega_1))$,
one finds that the integral including $1/(1+\mfa(\omega_1))$
is equal to zero since the integrand is  antisymmetric with respect to
$\omega_1$ and $\omega_2$. After shifting the variables
 $\zeta_j\to\zeta_j+\pi \i/4$ and $\omega_j\to\omega_j+\pi \i/4$, 
one obtains
\begin{align}
\bra  c_1 & c_{m+1}^{\dagger}\ket
=-8 (-1)^m \int_{\Gamma_{-\frac{\pi}{4}\i}} 
 \frac{\d \zeta}{2\pi \i}\[\frac{\sh(\zeta-\frac{\pi}{4}\i)}
                                {\sh(\zeta+\frac{\pi}{4}\i)}
                         \]^m 
 \frac{1}{\sh(2(\zeta+\frac{\pi}4\i))}
 \int_{\mathcal{C'}}\frac{\d \omega_1}{2\pi \i}
 \int_{\mathcal{C'}}\frac{\d \omega_2}{2\pi \i}
\frac{1}{1+\mfa(\omega_2+\frac{\pi}{4}\i)}             \nn \\
&
\times
\frac{\ch(\omega_1-\zeta)\ch(\omega_2-\zeta)}
     {\ch(\omega_1-\omega_2)}
     \[
       \frac{1}{\sh(2(\omega_1-\zeta))\sh(2(\omega_2+\frac{\pi}4\i))}
      -\frac{1}{\sh(2(\omega_2-\zeta))\sh(2(\omega_1+\frac{\pi}4\i))}
     \],  \nn
\end{align}
where $\Gamma_{-\pi \i/4}$ surrounds the point $\zeta=-\pi \i/4$;
$\mathcal{C}'=-\mathcal{C}_0\cup\mathcal{C}_{-\pi \i/2}$;
$\mathcal{C}_0=[-\infty,\infty]$;
$\mathcal{C}_{-\pi \i/2}=[-\pi \i/2-\infty,-\pi \i/2+\infty]$.
The integral with respect to $\omega_1$ can be easily
evaluated via the residue theorem applied to the poles
at $\omega=-\pi \i/4$ and $\zeta$. Then taking into account the
pole outside the contour $\Gamma_{-\pi \i/4}$ i.e. at the
point $\zeta=\omega_2$, we compute the integral
with respect to $\zeta$. It reads
\begin{align}
\bra c_1c_{m+1}^{\dagger}\ket=& 2 (-1)^m
  \int_{\mathcal{C}'}\frac{\d \omega}{2\pi \i}
                      \frac{1}{1+\mfa(\omega+\frac{\pi}4\i)}
                         \[
                           \frac{\sh(\omega-\frac{\pi}{4}\i)}
                                {\sh(\omega+\frac{\pi}{4}\i)}
                         \]^m 
                      \frac{1}{\sh(2(\omega+\frac{\pi}4\i))} \nn \\ 
  =&-2 (-1)^m
      \int_{-\infty}^{\infty}\frac{\d \omega}{2\pi \i}
                      \frac{1}{1+\mfa(\omega+\frac{\pi}4\i)}
                         \[
                           \frac{\sh(\omega-\frac{\pi}{4}\i)}
                                {\sh(\omega+\frac{\pi}{4}\i)}
                         \]^m 
                      \frac{1}{\sh(2(\omega+\frac{\pi}4\i))} \nn \\
&+2 (-1)^m
  \int_{-\infty}^{\infty}\frac{\d \omega}{2\pi \i}
                      \frac{1}{1+\mfa(\omega-\frac{\pi}4\i)}
                         \[
                          - \frac{\sh(\omega+\frac{\pi}{4}\i)}
                                 {\sh(\omega-\frac{\pi}{4}\i)}
                         \]^m 
                      \frac{1}{\sh(2(\omega-\frac{\pi}4\i))}.   \nn
\end{align}
Changing the variable $\cosh (2\omega)=1/\cos p$ ($p\in[-\pi/2,\pi/2]$)
for the first term in the second equality, and  $\cosh (2\omega)=-1/\cos p$ 
($p\in [-\pi,-\pi/2]\cup[\pi/2,\pi]$) for the second term, we obtain
\begin{equation}
\bra c_{1}c_{m+1}^{\dagger} \ket =
\frac{1}{2\pi}\int_{-\pi}^{\pi}\frac{\d p  \e^{\i m p}}
                                  {1+\exp\[-\frac{\mu_{\rm c}}{T}-\frac{2t}{T}\cos p \]}
=
-\frac{1}{2\pi}\int_{-\pi}^{\pi}\frac{\d p  \e^{\i m p}}
                                  {1+\exp\[\frac{\mu_{\rm c}}{T}+\frac{2t}{T}\cos p \]}.
\label{free-green}
\end{equation}
We note that the above expression reproduces the well-known result
(see \cite{FW} for example). 
Of course, \eqref{free-green} can also be derived by starting from 
\eqref{mult-green2}.

%
\section{Discussion}
%
We have derived the multiple integral representation for the 
equal-time one-particle Green's function of the spinless fermion 
model at finite temperature. Unfortunately, the explicit evaluation
of the multiple integrals still remains a difficult task, except
for some special cases considered here. Nevertheless, we believe that
the method provided in this paper should be useful for the
future study of the correlation functions of the fermionic systems.

For instance, from \eqref{correlation2} one sees that the long-distance behavior 
of the two-point correlation functions can be calculated by taking the 
ratio between the largest and the subleading eigenvalues of the quantum 
transfer matrix. For the one-particle Green's function, it reads
(up to the sign)
\begin{equation}
\bra c_1 c_{m+1} ^\dagger \ket \sim 
   2  A_0 \cos(k_{\rm F}(m-1)) \exp\[-\frac{m-1}{\xi}\], \nn
\end{equation}
with
\begin{equation}
  k_{\rm F}=\Im \[\ln \frac{\Lambda_1(0)}{\Lambda_0(0)}\],
  \quad
 -\frac{1}{ \xi}=\Re\[\ln \frac{\Lambda_1(0)}{\Lambda_0(0)}      \], \quad
A_0=\left|\frac{\bra \Psi_0| B(0) |\Psi_1 \ket}{\Lambda_0(0)}\right|^2,  \nn
\end{equation}
where $\Lambda_1(0)$ is the leading eigenvalue for the sector
$M=N/2-1$ (see \eqref{bae} and \eqref{eigenvalue}) and
$|\Psi_1\ket$ is the corresponding (normalized) eigenvector.
In fact the finite temperature correlation length $\xi$ has already been
calculated in \cite{SSSU}. The evaluation of the
amplitude $A_0$ by using \eqref{action} and Proposition~\ref{scalar} is
quite important problem.

It is also interesting to extend our result to the time dependent
case. This is possible by combining the present method with
the solution of the quantum inverse scattering problem for 
the operator $c_j$. It is evidently worth while to  extract the long-distance and
long-time behavior of the correlation functions at any
finite temperatures.
%
\section*{Acknowledgments}
%
This work is partially supported by Grants-in-Aid for 
Young Scientists (B) No.~17740248,
Scientific Research (B) No.~18340112 and (C) No.~18540341 from
the Ministry of Education, Culture, Sports, Science and Technology 
of Japan.
%
%
%
\rnc{\theequation}{A.\arabic{equation}}\setcounter{equation}{0}
\section*{Appendix. Derivation of multiple integral \eqref{mult-inhom1}}
%
We  describe how the multiple integral representation 
\eqref{mult-inhom1} is derived. Applying the relation 
\eqref{action} to the term
$\bra \Omega|\prod_{j=1}^{n} C(\txi^+_j)\prod_{k=1}^{N/2+1-n}
                             C(\tlam_k^-) B(\xi_{m+1})$
in the r.h.s of \eqref{mult2}, we split $\Phi_N(\{\xi\})$ into four parts 
according to whether the arguments of the functions $a(x_a)$ and $d(x_d)$ 
appearing in the resulting equation are Bethe roots $\{\lambda \}$ or 
inhomogeneous parameters $\{\xi\}$:
\begin{align}
\Phi_N(\{\xi\})=&
F_1(\{x_a\}_{\in \{\lambda\}}|\{x_d\}_{\in\{\lambda\}})+
F_2(\{x_a\}_{\in \{\lambda\}}|\{x_d\}_{\in\{\xi\}}) \nn \\
&
\quad + F_3(\{x_a\}_{\in \{\xi\}}|\{x_d\}_{\in\{\lambda\}})+
F_4(\{x_a\}_{\in \{\xi\}}|\{x_d\}_{\in\{\xi \}}).
\label{division}
\end{align}
First we consider the function $F_1$ which can further be 
divided into two parts according to whether $\xi_1 \in \{\tlam^+\}$ 
or $\xi_1\in \{\tlam^-\}$:
$
F_1=F_{\xi_1\in\{\tlam^+\}}+F_{\xi_1\in\{\tlam^-\}},
$
where
\begin{align}
&F_{\xi_1\in\{\tlam^+\}}=(-1)^{\frac{N}{2}+1}
  \sum_{n=1}^{m-1}
  \sum_{\substack{\{\tlam\}=\{\tlam^+\}\cup\{\lambda^-\} \\
                  \{\txi\}=\{\txi^+\}\cup\{\txi^-\}\\
                  |\tlam^+|=|\txi^+|=n}}
  \sum_{l=1}^{\frac{N}{2}-n+1}
  \sum_{\substack{l'=1\\l'\neq l}}^{\frac{N}{2}-n+1}
 \frac{
  H_{n}^{(1)}(\{\lam^-\}|\{\txi^+\})  
  H_{n}^{(2)}(\{\lam^-\}|\{\txi^+\})      
      }{\bra \{\lam\}|\{\lam\} \ket \prod_{j=1}^{m+1}\Lambda_0(\xi_j)}   \nn \\
& \quad
  \times R_n(\{\txi^+\}|\{\txi^-\}|\{\tlam^+\}|\{\lambda^-\})  
 \bra \Omega |\prod_{\substack{j=1 \\ j\neq l,l'}}^{\frac{N}{2}-n+1}
      C(\lambda_j^-)\prod_{j=1}^nC(\txi_j^+)C(\xi_{m+1})
     \prod_{j=1}^{\frac{N}{2}}B(\lambda_j)|\Omega\ket         
\label{F1}
\end{align}
with
\begin{align}
H_{n}^{(1)}(\{\lam^-\}|\{\txi^+\})&=
        \frac{d(\lambda_l^-)\sh (\eta)
               \prod_{\substack{j=1 \\ j\neq l}}^{\frac{N}{2}-n+1}
                       f(\lambda_l^-,\lambda_j^-)
               \prod_{j=1}^n f(\lambda_l^-,\txi_j^+)}{\sh(\lambda_l^--\xi_{m+1})},   
                                                                         \nn \\
H_{n}^{(2)}(\{\lam^-\}|\{\txi^+\})&=
        \frac{a(\lambda_{l'}^-) \sh(\eta)
                 f(\xi_{m+1},\lambda_{l'}^-)
               \prod_{\substack{j=1 \\ j\neq l,l'}}^{\frac{N}{2}-n+1}
                       f(\lambda_j^-,\lambda_{l'}^-)
               \prod_{j=1}^n f(\txi_j^+,\lambda_{l'}^-)}
            {\sh(\xi_{m+1}-\lambda_{l'}^-+\eta)},          \nn
\end{align}
while $F_{\xi_1\in\{\tlam^-\}}$ is 
\begin{align}
&F_{\xi_1\in\{\tlam^-\}}=(-1)^{\frac{N}{2}+1}
  \sum_{n=0}^{m-1}
  \sum_{\substack{\{\tlam\}=\{\lambda^+\}\cup\{\tlam^-\} \\
                  \{\txi\}=\{\txi^+\}\cup\{\txi^-\}\\
                  |\lambda^+|=|\txi^+|=n}}
  \sum_{l=1}^{\frac{N}{2}-n}
  \sum_{\substack{l'=1\\l'\neq l}}^{\frac{N}{2}-n}
 \frac{
  H_{n}^{(1)}(\{\tlam^-\}|\{\txi^+\})  
  H_{n}^{(2)}(\{\tlam^-\}|\{\txi^+\}) 
       } 
      {\bra \{\lam\}|\{\lam\} \ket \prod_{j=1}^{m+1}\Lambda_0(\xi_j)}        \nn \\
&\quad \times
  R_n(\{\txi^+\}|\{\txi^-\}|\{\lambda^+\}|\{\tlam^-\})    
      \bra \Omega |\prod_{\substack{j=1 \\ j\neq l,l'}}^{\frac{N}{2}-n+1}
      C(\tlam_j^-)\prod_{j=1}^nC(\txi_j^+)C(\xi_{m+1})
     \prod_{j=1}^{\frac{N}{2}}B(\lambda_j)|\Omega\ket. \nn
\end{align}
Here $\{\lam^{\pm}\}$ denote $\{\lam^{\pm}\}=\{\tlam^{\pm}\}^{\setminus \xi_1}$.

Inserting the relations \eqref{eigenvalue}, \eqref{auxiliary}, 
\eqref{R}, \eqref{ratio} and $\mfa(\lam_j^+)=-1$ into \eqref{F1} and shifting
the variable $n\to n+1$, we have
\begin{align}
F_{\xi_1\in\{\tlam^+\}}=&
  \sum_{n=0}^{m-2}
  \sum_{\substack{\{\lam\}=\{\lam^+\}\cup\{\lam^-\} \\
                  \{\txi\}=\{\txi^+\}\cup\{\txi^-\}\\
                  |\lam^+|=|\txi^+|-1=n}}
  \sum_{l=1}^{\frac{N}{2}-n}
  \sum_{\substack{l'=1\\l'\neq l}}^{\frac{N}{2}-n}
  \frac{(-1)^{n}\mfa(\lam_l^-)}{\mfa'(\lam^-_l)\mfa'(\lam^-_{l'})
                                              \prod_{j=1}^{n}\mfa'(\lam_j^+)}  \nn \\
&
  \times\frac{\tY_{n}(\{\lam^+\}|\{\txi^+\})Z_{n}(\{\lam^+\}|\{\txi\}) 
   V_n^{+}(\{\lam^+\}|\{\txi^+\}) X_n(\{\lam^+\}|\{\txi^+\})}
   {\sh(\lam_l^--\lam^-_{l'}+\eta)(1+\mfa(\xi_1)) \prod_{j=1}^{m-n-2}(1+\mfa(\txi_j^-))},
\label{F1new}   
\end{align}
where the functions $\tY_{n}(\{\lam^+\}|\{\txi^+\})$,
$V_n^{\pm}(\{\lam^+\}|\{\txi^+\})$, $X_n(\{\lam^+\}|\{\txi^+\})$ and
$Z_{n}(\{\lam^+\}|\{\txi\})$ are defined as follows:
\begin{align}
\tY_{n}(\{\lam^+\}|\{\txi^+\})=&
    \frac{\prod_{j=1}^n\mfb_+(\lam_j^+)}{\prod_{j=1}^{n+1}
                      \mfb'_+ (\txi_j^+)}
    \frac{\prod_{j=1}^n \prod_{k=1}^{n+1}
              [\sh(\lam_j^+-\txi_k^+-\eta)\sh(\lam_j^+-\txi_k^+ +\eta)]}
         {\prod_{j,k=1}^{n+1}\sh(\txi_j^+-\txi_k^++\eta)
          \prod_{j,k=1}^n \sh(\lam_j^+-\lam_k^+-\eta)}  \nn \\
  &\times {\det}_{n+1} \tM_{j k} {\det}_{n+2} G(\widehat{\lam}_j,\widehat{\xi}_k)  \nn
\end{align}
with 
\begin{align}
&\mfb_{\pm}(\lam)=\prod_{j=1}^{m-1}\frac{\sh(\lam-\txi_j)}{\sh(\lam-\txi_j\pm\eta)}=
              \prod_{j=2}^{m}\frac{\sh(\lam-\xi_j)}{\sh(\lam-\xi_j\pm\eta)},\nn \\
&\tM_{jk}=\begin{cases}
           t(\txi_k^+,\lam_j^+)-t(\lam_j^+,\txi_k^+)
               \prod_{a=1}^n \frac{f(\lam_a^+,\lam_j^+)}{f(\lam_j^+,\lam_a^+)}
               \prod_{b=1}^{n+1}\frac{f(\lam_j^+,\txi_b^+)}{f(\txi_b^+,\lam_j^+)} 
                                   & \text{ for $j\le n$}             \\
           t(\txi_k^+,\xi_1)+\mfa(\xi_1)t(\xi_1,\txi_k^+)
               \prod_{a=1}^n \frac{f(\lam_a^+,\xi_1)}{f(\xi_1,\lam_a^+)}
               \prod_{b=1}^{n+1}\frac{f(\xi_1,\txi_b^+)}{f(\txi_b^+,\xi_1)} 
                                   & \text{ for $j= n+1$}    
        \end{cases},  \nn
\end{align}
and $G(\hlam,\hxi)$ is the solution of  the linear integral
equation \eqref{G-function}, where the variables 
$\{\hlam_j\}_{j=1}^{n+2}$ and $\{\hxi_j\}_{j=1}^{n+2}$ are, 
respectively, assigned as
\begin{equation}
(\hlam_1,\dots,\hlam_{n+2})=
   (\lam^+_1,\dots,\lam^+_{n},\lam^-_l,\lam^-_{l'}),
\quad
(\hxi_1,\dots,\hxi_{n+2})=
   (\txi^+_1,\dots,\txi^+_{n+1},\xi_{m+1});  \nn
\end{equation}
\begin{align}
&
V_n^{\pm}(\{\lam^+\}|\{\txi^+\})=\frac{\prod_{j=1}^n \sh(\lam_j^+-\xi_{m+1})
       \prod_{j=1}^{n+1} \sh(\txi_j^+-\xi_{1} \pm \eta)}
      {\prod_{j=1}^{n+1}\sh(\txi_j^+-\xi_{m+1})
       \prod_{j=1}^{n}\sh(\lam_j^+-\xi_{1}\pm \eta)},  \nn \\
&
X_n(\{\lam^+\}|\{\txi^+\})=\frac{\prod_{j=1}^{n+1} [\sh(\lam_l^--\txi_j^++\eta) 
                          \sh(\lam_{l'}^--\txi_j^+-\eta)]}
      {\prod_{j=1}^{n} [\sh(\lam_l^--\lam_j^++\eta) 
                          \sh(\lam_{l'}^--\lam_j^+-\eta)]}, \nn \\
&
Z_{n}(\{\lam^+\}|\{\txi\})=\prod_{j=1}^{|\txi^-|}
           \[1-\mfa(\txi^-_j)\prod_{k=1}^{n} 
                    \frac{f(\lam_k^+,\txi_j^-)}
                         {f(\txi_j^-,\lam_k^+)}
                \prod_{k=1}^{n+1} 
                    \frac{f(\txi_j^-,\txi_k^+)}
                         {f(\txi_k^+,\txi_j^-)}
           \],                      \nn
\end{align}
where $|\txi^-|=m-n-2$.
Similarly, we obtain
\begin{align}
F_{\xi_1\in\{\tlam^-\}}=&
  \sum_{n=0}^{m-1}
  \sum_{\substack{\{\lam\}=\{\lam^+\}\cup\{\lam^-\} \\
                  \{\txi\}=\{\txi^+\}\cup\{\txi^-\}\\
                  |\lam^+|=|\txi^+|=n}}
  \sum_{l=1}^{\frac{N}{2}-n}
  \sum_{\substack{l'=1\\l'\neq l}}^{\frac{N}{2}-n}
  \frac{\mfa(\lam_l^-)}{\mfa'(\lam_l)\mfa'(\lam_{l'}) \prod_{j=1}^{n}\mfa'(\lam_j^+)}      \nn \\
&
  \times\frac{\bY_{n}(\{\lam^+\}|\{\bxi^+\})Z_{n}(\{\lam^+\}|\{\bxi\})
              V_n^+(\{\lam^+\}|\{\bxi^+\}) X_n(\{\lam^+\}|\{\bxi^+\})}
   {\sh(\lam_l^--\lam^-_{l'}+\eta)\prod_{j=1}^{m-n-1}(1+\mfa(\bxi_j^-))},
\label{F2new}   
\end{align}
where $(\bxi^+_1,\dots,\bxi^+_{n+1})=(\xi_1,\txi_1^{+},\dots,\txi_{n}^+)$,
$\{\bxi_j^-\}_{j=1}^{m-n-1}=\{\txi_j^-\}_{j=1}^{m-n-1}$,
$\{\bxi\}=\{\bxi^+\}\cup \{\bxi^-\}$ and
\begin{align}
&\bY_{n}(\{\lam^+\}|\{\txi^+\})=
    \frac{\prod_{j=1}^n \mfb_+(\lam_j^+)}{\mfb_+(\xi_1)\prod_{j=1}^n \mfb'_+ (\txi_j^+)}
    \frac{\prod_{j=1}^n \prod_{k=1}^{n+1}
              [\sh(\lam_j^+-\bxi_k^+-\eta)\sh(\lam_j^+-\bxi_k^+ +\eta)]}
         {\prod_{j,k=1}^{n+1}\sh(\bxi_j^+-\bxi_k^++\eta)
          \prod_{j,k=1}^n \sh(\lam_j^+-\lam_k^+-\eta)}  \nn \\
  &\qquad\qquad\times {\det}_{n} \bM_{j k} {\det}_{n+2} G(\widehat{\lam}_j,\ckxi_k), 
  \quad (\ckxi_1,\dots,\ckxi_{n+2})=(\bxi^+_1,\dots,\bxi^+_{n+1},\xi_{m+1}), \nn \\
&\bM_{jk}= t(\txi_k^+,\lam_j^+)-t(\lam_j^+,\txi_k^+)\prod_{a=1}^n 
                      \frac{f(\lam_a^+,\lam_j^+)}{f(\lam_j^+,\lam_a^+)}
           \prod_{b=1}^{n+1}\frac{f(\lam_j^+,\bxi_b^+)}{f(\bxi_b^+,\lam_j^+)}.
\nn
\end{align}

Now we express the sums over partitions  in \eqref{F1new} and
\eqref{F2new} as multiple integrals. 
To this end, we would like to  introduce the following useful formula. 
Let $f(\omega_1,\dots,\omega_n)$ be a function which is
analytic on and inside the contour 
$\mathcal{C}$, symmetric with respect to $\{\omega_j\}_{j=1}^n$,
and zero when any two of its variables are the same. The
poles of the function $1/(1+\mfa(\omega))$ inside $\mathcal{C}$
are simple poles at $\omega=\lambda_j$ with residues $1/\mfa'(\lambda_j)$, 
where $\{\lambda_j\}_{j=1}^{N/2}$ are the Bethe roots 
characterizing the largest eigenvalue of the quantum transfer matrix.
Hence one has
\begin{equation}
\frac{1}{n!}\int_{\mathcal{C}^n}\prod_{j=1}^n
       \frac{\d\omega_j}{2\pi\i(1+\mfa(\omega_j))}
       f(\omega_1,\dots,\omega_n)=
   \sum_{\substack{\{\lambda\}=\{\lambda^+\}\cup\{\lambda^-\} \\
                   |\lambda^+|=n}} 
    \frac{f(\lambda_1^+,\dots,\lambda_n^+)}{\prod_{j=1}^n \mfa'(\lambda_j^+)}.
\label{sum2int}
\end{equation}
The relation similar to the above  also holds for $\mfab$.

First we apply \eqref{sum2int} to the partition for 
the set of the Bethe roots $\{\lam\}$ in \eqref{F1new}.
We see that the summand in \eqref{F1new} has simple poles inside
$\mathcal{C}$ at $\hlam_j=\txi^+_k$. Since the inhomogeneous 
parameters $\{\xi\}$ can be chosen arbitrary values,
we choose $\{\xi \}$ such that the two  sets of parameters $\{\xi\}$
and $\{\lam\}$ are distinguishable. Then there exists
a simple closed contour surrounding the Bethe roots $\{\lam\}$
but excluding $\{\xi\}$. Let $\mathcal{C}-\Gamma$ be such a
contour, where $\Gamma$ encircles $\{\xi\}$.
Applying \eqref{sum2int} into \eqref{F1new}, one has
\begin{align}
F_{\xi_1\in\{\tlam^+\}}=&
  \sum_{n=0}^{m-2}
  \sum_{\substack{\{\txi\}=\{\txi^+\}\cup\{\txi^-\}\\
                  |\txi^+|=n+1}}
  \sum_{l=1}^{\frac{N}{2}-n}
  \sum_{\substack{l'=1\\l'\neq l}}^{\frac{N}{2}-n}
  \frac{1}{n!}
  \int_{(\mathcal{C}-\Gamma)^n}
    \prod_{j=1}^n\[\frac{\d \omega_j}
                             {2\pi \i (1+\mfa(\omega_j))}
                 \]
  \frac{(-1)^{n}\mfa(\lam_l^-)}{\mfa'(\lam^-_l)
                                  \mfa'(\lam^-_{l'})}  \nn \\
&
  \times\frac{\tY_{n}(\{\omega \}|\{\txi^+\})Z_{n}(\{\omega\}|\{\txi\})
 V_n^+(\{\omega\}|\{\txi^+\}) X_n(\{\omega\}|\{\txi^+\})}
   { \sh(\lam_l^--\lam^-_{l'}+\eta)(1+\mfa(\xi_1))\prod_{j=1}^{m-n-2}(1+\mfa(\txi_j^-))}.
\label{F1int}   
\end{align}
By dividing the integrals, we transform the integrals along the contour 
$\mathcal{C}-\Gamma$  to those along the canonical contour $\mathcal{C}$:
\begin{equation}
\int_{(\mathcal{C}-\Gamma)^n}
\prod_{j=1}^n \frac{\d \omega_j}{2\pi\i}
\longrightarrow
\sum_{k=0}^{n}(-1)^k\binom{n}{k}
\int_{\mathcal{C}^{n-k}}
\prod_{j=1}^{n-k} \frac{\d\omega_j}{2\pi \i}
\int_{\Gamma^k}\prod_{j=1}^{k}  \frac{\d \omega_{n-k+j}}{2\pi \i},
\label{seperation}
\end{equation}
where we have used the fact that the integrand in \eqref{F1int}
is symmetric with respect to $\{\omega\}$.
Noting that, inside $\Gamma$, the integrand has simple poles at
$\omega_j=\txi_k^+$, one can explicitly
calculate the integrals over $\Gamma$:
\begin{align}
&
\int_{\Gamma^k}
    \prod_{j=1}^k\[\frac{\d \omega_{n-k+j}}
                             {2\pi \i (1+\mfa(\omega_{n-k+j}))}
                 \]
\tY_{n}(\{\omega \}|\{\txi^+\})Z_{n}(\{\omega\}|\{\txi\})
 V_n^+(\{\omega\}|\{\txi^+\}) X_n(\{\omega\}|\{\txi^+\})  \nn \\
&\,\,=k! \sum_{\substack{\{\txi^{++}\}\cup\{\txi^{+-}\}=\{\txi^+\} \\
                                                     |\txi^{+-}|=k}}
        \tY_{n-k}(\{\omega_j\}_{j=1}^{n-k}|\{\txi^{++}\})   
        V^+_{n-k}(\{\omega_j\}_{j=1}^{n-k}|\{\txi^{++}\})
        X_{n-k}(\{\omega_j\}_{j=1}^{n-k}|\{\txi^{++}\})     \nn \\
& \qquad \times \prod_{j=1}^k\frac{1}{1+\mfa(\txi_j^{+-})}
                \prod_{j=1}^{m-n-2}
                   \[
                     1-\mfa(\txi_j^-)\prod_{a=1}^{n-k} \frac{f(\omega_a,\txi_j^-)}
                                                            {f(\txi_j^-,\omega_a)}
                                     \prod_{b=1}^{n-k+1}\frac{f(\txi_j^-,\txi_b^{++})}
                                                             {f(\txi_b^{++},\txi_j^-)}
                   \]                                                             \nn \\
& \qquad \times  \prod_{j=1}^{k}
                   \[
                     1+\prod_{a=1}^{n-k} \frac{f(\omega_a,\txi_j^{+-})}
                                              {f(\txi_j^{+-},\omega_a)}
                                     \prod_{b=1}^{n-k+1}\frac{f(\txi_j^{+-},\txi_b^{++})}
                                                             {f(\txi_b^{++},\txi_j^{+-})}
                   \].
\label{F1divide}   
\end{align}
By inserting \eqref{F1divide} via \eqref{seperation} into \eqref{F1int}, 
the integrals on the contour $\mathcal{C}-\Gamma$ can be transformed to 
those on the canonical contour $\mathcal{C}$.

The remaining task is the calculation of the sums over the 
partition of inhomogeneous parameters $\{\xi\}$. Resumming
them by using the formula as in \cite{GKS04}
\begin{equation}
\sum_{k=0}^{|x|}(-1)^k  \sum_{\substack{\{x^+\}\cup\{x^-\}=\{x\}  \\
                                       |x^+|=k}}
                      \prod_{j=1}^{|x^-|}\[1+\kappa f(x_j^-) g(x_j^-)\]
                      \prod_{j=1}^{|x^+|}\[1-\kappa          g(x_j^+)\]  
          =\kappa^{|x|}\prod_{j=1}^{|x|}\[ g(x_j)(1+f(x_j)) \],   \nn
\end{equation} 
and further expressing the sum over $\lam_l^-$ (respectively
$\lam^-_{l'}$) as the integral over $\omega_{n+1}$ (respectively
$\omega_{n+2}$) by \eqref{sum2int}, one has
\begin{align}
F_{\xi_1\in\{\tlam^+\}}&=
  \sum_{n=0}^{m-2}
  \sum_{\substack{\{\txi\}=\{\txi^+\}\cup\{\txi^-\}\\
                  |\txi^+|=n+1}}
  \frac{(-1)^{m}}{n!}
  \int_{\mathcal{C}^n}
    \prod_{j=1}^n\[\frac{\d \omega_j \mfb_-(\omega_j)}
                             {2\pi \i (1+\mfa(\omega_j))}
                 \]
  \int_{\mathcal{C}-\Gamma}
       \frac{\d \omega_{n+1}}{2\pi \i (1+\mfab(\omega_{n+1}))}  \nn \\
&
  \times 
   \int_{\mathcal{C}-\Gamma}
       \frac{\d \omega_{n+2}}{2\pi \i (1+\mfa(\omega_{n+2}))} 
  \frac{V_n^+(\{\omega\}|\{\txi^+\}) W^-_n(\{\omega\}|\{\txi^+\}) 
                                           X_n(\{\omega\}|\{\txi^+\})}
   { \sh(\omega_{n+1}-\omega_{n+2}+\eta)\prod_{j=1}^{n+1} 
                                \mfb'_-(\txi_j^+)(1+\mfa(\xi_1))} \nn \\
&
  \times {\det}_{n+1}M_{jk}^{(1)}(\{\omega\}|\{\txi^+\})
         {\det}_{n+2}[G(\omega_j,\txi_1^+),\dots,G(\omega_j,\txi_{n+1}^+),G(\omega_j,\xi_{m+1})],
\label{F1resum}   
\end{align}
where $W^-_n(\{\omega\}|\{\txi^+\})$ and the $(n+1) \times (n+1)$ matrix 
$M^{(1)}_{jk}(\{\omega\}|\{\txi^+\})$ are, respectively, defined by
\begin{align}
&W^{\pm}_n(\{\omega\}|\{\txi^+\})=
            \frac{\prod_{j=1}^{n}\prod_{k=1}^{n+1}\sh(\omega_j-\txi^+_k \pm \eta)
                                                  \sh(\txi^+_k-\omega_{j}\pm \eta)}
                 {\prod_{j,k=1}^{n}\sh(\omega_{j}-\omega_{k}\pm \eta)
                  \prod_{j,k=1}^{n+1}\sh(\txi^+_j-\txi^+_k\pm \eta)},      \nn \\
&M_{jk}^{(1)}(\{\omega\}|\{\txi^+\}) \nn \\
& \quad   =\begin{cases}
           t(\omega_j,\txi_k^+)+t(\txi_k^+,\omega_j)
               \prod_{a=1}^{n}\frac{\sh(\omega_a-\omega_j-\eta)}
                                   {\sh(\omega_j-\omega_a-\eta)} 
               \prod_{b=1}^{n+1} \frac{\sh(\omega_j-\txi_b^+-\eta)}
                                      {\sh(\txi_b^+-\omega_j -\eta)}
                                   & \text{ for $j\le n$}             \\
           t(\txi_k^+,\xi_1)+\mfa(\xi_1)t(\xi_1,\txi_k^+)
               \prod_{a=1}^n \frac{\sh(\omega_a-\xi_1+\eta)}{\sh(\omega_a-\xi_1-\eta)}
               \prod_{b=1}^{n+1}\frac{\sh(\txi_b^+-\xi_1-\eta)}{\sh(\txi_b^+-\xi_1+\eta)} 
                                   & \text{ for $j= n+1$}    
        \end{cases}.   \nn
\end{align}
The integrand of \eqref{F1resum} is a symmetric function 
with respect to $\{\txi^+\}$ and vanishes when any two of them
are the same. Thanks to this together with the fact
that $1/\mfb(\omega)$ has simple poles at $\omega=\txi_k$,
we can directly apply \eqref{sum2int} to \eqref{F1resum}.
Thus we arrive at
\begin{align}
&F_{\xi_1\in\{\tlam^+\}}=
  \sum_{n=0}^{m-2}
  \frac{(-1)^{m}}{n!(n+1)!}
  \int_{\tGamma^{n+1}}
    \prod_{j=1}^{n+1}\[\frac{\d \zeta_j }{2\pi \i \mfb_-(\zeta_j)} \]
  \int_{\mathcal{C}^n}
    \prod_{j=1}^n\[\frac{\d \omega_j \mfb_-(\omega_j)}
                             {2\pi \i (1+\mfa(\omega_j))}
                 \]                                             \nn \\
& \quad \times
  \int_{\mathcal{C}-\Gamma}
       \frac{\d \omega_{n+1}}{2\pi \i (1+\mfab(\omega_{n+1}))}  
   \int_{\mathcal{C}-\Gamma}
       \frac{\d \omega_{n+2}}{2\pi \i (1+\mfa(\omega_{n+2}))} 
  \frac{V_n^+(\{\omega\}|\{\zeta\}) W^-_n(\{\omega\}|\{\zeta\})X_n(\{\omega\}|\{\zeta\})}
   { \sh(\omega_{n+1}-\omega_{n+2}+\eta)(1+\mfa(\xi_1))} \nn \\
&
  \quad \times 
         {\det}_{n+1}M_{jk}^{(1)}(\{\omega\}|\{\zeta\})
         {\det}_{n+2}[G(\omega_j,\zeta_1),\dots,G(\omega_j,\zeta_{n+1}),
                      G(\omega_j,\xi_{m+1})],
\label{F1mult}   
\end{align}
where $\tGamma=\Gamma-\Gamma_{\xi_1}$; $\Gamma_{\xi_1}$ surrounds
the point $\xi_1$ but excludes $\{\txi\}$.

Almost the same method is applied to $F_{\xi_1\in\{\tlam^-\}}$ \eqref{F2new}
by considering the integrals over the contour 
$\tC=\mathcal{C}-\Gamma_{\xi_1}$ 
instead of $\mathcal{C}$.
Utilizing the transformation \eqref{sum2int}, and  resumming the
resulting equation as in the case of $F_{\xi_1\in\{\tlam^+\}}$,
one may have
\begin{align}
&F_{\xi_1\in\{\tlam^-\}}=
  \sum_{n=0}^{m-1}
  \sum_{\substack{\{\txi\}=\{\txi^+\}\cup\{\txi^-\}\\
                  |\txi^+|=n}}
  \frac{(-1)^{m-n-1}}{n!}
  \int_{\mathcal{\tC}^n}
    \prod_{j=1}^n\[\frac{\d \omega_j \mfb_-(\omega_j)}
                             {2\pi \i (1+\mfa(\omega_j))}
                 \]
  \int_{\mathcal{C}-\Gamma}
       \frac{\d \omega_{n+1}}{2\pi \i (1+\mfab(\omega_{n+1}))}  \nn \\
& \quad
  \times 
   \int_{\mathcal{C}-\Gamma}
       \frac{\d \omega_{n+2}}{2\pi \i (1+\mfa(\omega_{n+2}))} 
  \frac{U_n(\{\omega\}|\{\bxi^+\})V_n^+(\{\omega\}|\{\bxi^+\}) 
        W^-_n(\{\omega\}|\{\bxi^+\})X_n(\{\omega\}|\{\bxi^+\})}
   { \sh(\omega_{n+1}-\omega_{n+2}+\eta)\mfb_-(\xi_1)\prod_{j=1}^{n} 
                                \mfb'_-(\txi_j^+)}  \nn \\
& \quad
  \times {\det}_{n}\hM_{jk}(\{\omega\}|\{\txi^+\})
         {\det}_{n+2}[G(\omega_j,\xi_1),G(\omega_j,\txi_{1}^+),\dots,
                      G(\omega_j,\txi_{n}^+),G(\omega_j,\xi_{m+1})],
\label{F2resum}   
\end{align}
where the function $U_n(\{\omega\}|\{\bxi^+\})$ and $n \times n$ matrix
$\hM_{jk}(\{\omega\}|\{\txi^+\})$ are, respectively, written as
\begin{align}
& U_n(\{\omega\}|\{\bxi^+\})=
    \prod_{j=1}^{n}\frac{\sh(\omega_j-\xi_1+\eta)}{\sh(\omega_j-\xi_1-\eta)}
    \prod_{k=1}^{n+1}\frac{\sh(\bxi_k^+-\xi_1-\eta)}{\sh(\bxi_k^+-\xi_1+\eta)}, \nn \\
& \hM_{jk}(\{\omega\}|\{\txi^+\})=
    t(\omega_j,\txi_k^+)+t(\txi_k^+,\omega_j)
    \prod_{a=1}^{n}
         \frac{\sh(\omega_a-\omega_j-\eta)}{\sh(\omega_j-\omega_a-\eta)}
    \prod_{b=1}^{n+1}
    \frac{\sh(\omega_j-\bxi_b^+-\eta)}{\sh(\bxi_b^+-\omega_j-\eta)}. \nn
\end{align}
Applying again the formula \eqref{seperation} to the integration
over $\tC=\mathcal{C}-\Gamma_{\xi_1}$, and noting that the sum over $k$
in \eqref{seperation} is restricted to $k=0$ and $k=1$, we divide
$F_{\xi_1\in\{\tlam^-\}}$ in \eqref{F2resum} into the following two
parts:
$
F_{\xi_1\in\{\tlam^-\}}=F_{\xi_1\in\{\tlam^-\}}^{(0)}+
                        F_{\xi_1\in\{\tlam^-\}}^{(1)},
$
where $F_{\xi_1\in\{\tlam^-\}}^{(0)}$ is given by simply changing the
contour $\tC \to\mathcal{C}$ in \eqref{F2resum}, while 
$F_{\xi_1\in\{\tlam^-\}}^{(1)}$ is written as
\begin{align}
&F_{\xi_1\in\{\tlam^-\}}^{(1)}=
  \sum_{n=0}^{m-2}
  \frac{(-1)^{m}}{n!(n+1)!}
  \int_{\tGamma^{n+1}}
    \prod_{j=1}^{n+1}\[\frac{\d \zeta_j }{2\pi \i \mfb_-(\zeta_j)} \]
  \int_{\mathcal{C}^n}
    \prod_{j=1}^n\[\frac{\d \omega_j \mfb_-(\omega_j)}
                             {2\pi \i (1+\mfa(\omega_j))}
                 \]                                             \nn \\
& \quad \times
  \int_{\mathcal{C}-\Gamma}
       \frac{\d \omega_{n+1}}{2\pi \i (1+\mfab(\omega_{n+1}))}  
   \int_{\mathcal{C}-\Gamma}
       \frac{\d \omega_{n+2}}{2\pi \i (1+\mfa(\omega_{n+2}))} 
  \frac{V_n^+(\{\omega\}|\{\zeta\}) W^-_n(\{\omega\}|\{\zeta\})X_n(\{\omega\}|\{\zeta\})}
   { \sh(\omega_{n+1}-\omega_{n+2}+\eta)(1+\mfa(\xi_1))} \nn \\
&
  \quad \times 
         {\det}_{n+1}M_{jk}^{(2)}(\{\omega\}|\{\zeta\})
         {\det}_{n+2}[G(\omega_j,\zeta_1),\dots,G(\omega_j,\zeta_{n+1}),
                      G(\omega_j,\xi_{m+1})].
\label{F21mult}   
\end{align}
Note here that we have shifted the variable $n\to n+1$ and converted
the sum over the partition for $\{\txi\}$ into the integrals
over $\tGamma$. The $(n+1)\times(n+1)$ matrix $M_{jk}^{(2)}(\{\omega\}|\{\zeta\}) $
is defined as
\begin{align}
&M_{jk}^{(2)}(\{\omega\}|\{\zeta\}) \nn \\
& \quad   =\begin{cases}
           t(\omega_j,\zeta_k)+t(\zeta_k,\omega_j)
               \prod_{a=1}^{n}\frac{\sh(\omega_a-\omega_j-\eta)}
                                   {\sh(\omega_j-\omega_a-\eta)} 
               \prod_{b=1}^{n+1} \frac{\sh(\omega_j-\zeta_b-\eta)}
                                      {\sh(\zeta_b-\omega_j -\eta)}
                                   & \text{ for $j\le n$}             \\
          - t(\zeta_k,\xi_1)+t(\xi_1,\zeta_k)
               \prod_{a=1}^n \frac{\sh(\omega_a-\xi_1+\eta)}{\sh(\omega_a-\xi_1-\eta)}
               \prod_{b=1}^{n+1}\frac{\sh(\zeta_b-\xi_1-\eta)}{\sh(\zeta_b-\xi_1+\eta)} 
                                   & \text{ for $j= n+1$}    
        \end{cases}.  \nn
\end{align}

In the next step, we would like to consider the sum 
$F_1=F_{\xi_1\in\{\tlam^+\}}+F_{\xi_1\in\{\tlam^-\}}$
and  combine the three multiple integrals into one.
First we deal with the sum $F_{\xi_1\in\{\tlam^+\}}+F_{\xi_1\in\{\tlam^-\}}^{(1)}$.
From \eqref{F1mult} and \eqref{F21mult}, it immediately follows that
\begin{align}
& F_{\xi_1\in\{\tlam^+\}}+F_{\xi_1\in\{\tlam^-\}}^{(1)}=
  \sum_{n=0}^{m-2}
  \frac{(-1)^{m}}{n!(n+1)!}
  \int_{\tGamma^{n+1}}
    \prod_{j=1}^{n+1}\[\frac{\d \zeta_j }{2\pi \i \mfb_-(\zeta_j)} \]
  \int_{\mathcal{C}^n}
    \prod_{j=1}^n\[\frac{\d \omega_j \mfb_-(\omega_j)}
                             {2\pi \i (1+\mfa(\omega_j))}
                 \]                                             \nn \\
& \quad \times
  \int_{\mathcal{C}-\Gamma}
       \frac{\d \omega_{n+1}}{2\pi \i (1+\mfab(\omega_{n+1}))}  
   \int_{\mathcal{C}-\Gamma}
       \frac{\d \omega_{n+2}}{2\pi \i (1+\mfa(\omega_{n+2}))} 
  \frac{V_n^-(\{\omega\}|\{\zeta\}) W^-_n(\{\omega\}|\{\zeta\}) X_n(\{\omega\}|\{\zeta\})}
   { \sh(\omega_{n+1}-\omega_{n+2}+\eta)} \nn \\
&
  \quad \times
         {\det}_{n+1}M_{jk}^{-}(\{\omega\}|\{\zeta\})
         {\det}_{n+2}[G(\omega_j,\zeta_1),\dots,G(\omega_j,\zeta_{n+1}),
                      G(\omega_j,\xi_{m+1})],  \nn
\end{align}
where the elements of the $(n+1)\times(n+1)$ matrix $M^{-}(\{\omega\}|\{\zeta\})$
are given by
\begin{align}
&M_{jk}^{-} =\begin{cases}
           t(\omega_j,\zeta_k)+t(\zeta_k,\omega_j)
               \prod_{a=1}^{n}\frac{\sh(\omega_a-\omega_j-\eta)}
                                   {\sh(\omega_j-\omega_a-\eta)} 
               \prod_{b=1}^{n+1} \frac{\sh(\omega_j-\zeta_b-\eta)}
                                      {\sh(\zeta_b-\omega_j -\eta)}
                                   & \text{ for $j\le n$}             \\
        t(\xi_1,\zeta_k)
                                   & \text{ for $j= n+1$}    
        \end{cases}.   \nn
\end{align}
Changing the contour $\tGamma\to\Gamma$ and combining it with 
$ F_{\xi_1\in\{\tlam^-\}}^{(0)}$, we obtain
\begin{align}
& F_1=
  \sum_{n=0}^{m-1}
  \frac{(-1)^{m}}{n!(n+1)!}
  \int_{\Gamma^{n+1}}
    \prod_{j=1}^{n+1}\[\frac{\d \zeta_j }{2\pi \i \mfb_-(\zeta_j)} \]
  \int_{\mathcal{C}^n}
    \prod_{j=1}^n\[\frac{\d \omega_j \mfb_-(\omega_j)}
                             {2\pi \i (1+\mfa(\omega_j))}
                 \]                                             \nn \\
& \quad \times
  \int_{\mathcal{C}-\Gamma}
       \frac{\d \omega_{n+1}}{2\pi \i (1+\mfab(\omega_{n+1}))}  
   \int_{\mathcal{C}-\Gamma}
       \frac{\d \omega_{n+2}}{2\pi \i (1+\mfa(\omega_{n+2}))} 
  \frac{V_n^-(\{\omega\}|\{\zeta\}) W^-_n(\{\omega\}|\{\zeta\}) X_n(\{\omega\}|\{\zeta\})}
   { \sh(\omega_{n+1}-\omega_{n+2}+\eta)} \nn \\
&
  \quad \times
         {\det}_{n+1}M_{jk}^{-}(\{\omega\}|\{\zeta\})
         {\det}_{n+2}[G(\omega_j,\zeta_1),\dots,G(\omega_j,\zeta_{n+1}),G(\omega_j,\xi_{m+1})].
\label{Fmult}   
\end{align}
The remaining contribution $F_2+F_3+F_4$ in \eqref{division} can be
absorbed into \eqref{Fmult} by changing the integration contours 
for $\omega_{n+1}$ and $\omega_{n+2}$ as $\mathcal{C}-\Gamma \to 
\mathcal{C}$. We thus finally arrive at
\begin{align}
&\Phi_N(\{\xi\})=
\sum_{n=0}^{m-1}
  \frac{(-1)^{m}}{n!(n+1)!}
  \int_{\Gamma^{n+1}}
    \prod_{j=1}^{n+1}\[\frac{\d \zeta_j }{2\pi \i \mfb_-(\zeta_j)} \]
  \int_{\mathcal{C}^n}
    \prod_{j=1}^n\[\frac{\d \omega_j \mfb_-(\omega_j)}
                             {2\pi \i (1+\mfa(\omega_j))}
                 \]                                             \nn \\
& \quad \times
  \int_{\mathcal{C}}
       \frac{\d \omega_{n+1}}{2\pi \i (1+\mfab(\omega_{n+1}))}  
   \int_{\mathcal{C}}
       \frac{\d \omega_{n+2}}{2\pi \i (1+\mfa(\omega_{n+2}))} 
  \frac{V_n^-(\{\omega\}|\{\zeta\}) W^-_n(\{\omega\}|\{\zeta\}) X_n(\{\omega\}|\{\zeta\})}
   { \sh(\omega_{n+1}-\omega_{n+2}+\eta)} \nn \\
&
  \quad \times
         {\det}_{n+1}M_{jk}^{-}(\{\omega\}|\{\zeta\})
         {\det}_{n+2}[G(\omega_j,\zeta_1),\dots,G(\omega_j,\zeta_{n+1}),
                      G(\omega_j,\xi_{m+1})].   \nn
\end{align}
%

\end{document}